\pdfoutput=1 

\documentclass[%
 reprint,
nofootinbib,
 amsmath,amssymb,
 aps,
]{revtex4-1}

\usepackage{preprintcover}  
\PreprintCoverPaperTitle{Measurement of $W\gamma$ and $Z\gamma$ production cross sections \\ in $pp$ collisions at $\sqrt{s}=$ 7 TeV and limits on anomalous \\ triple gauge couplings with the ATLAS detector}  
\PreprintIdNumber{CERN-PH-EP-2012-059}  
\PreprintCoverAbstract{  This Letter presents measurements of $l^{\pm}\nu\gamma$ and
  $l^{+}$$l^{-}$$\gamma$ ($l=e$, $\mu$) production in 1.02 fb$^{-1}$ of $pp$ collision data
  recorded at $\sqrt{s}$~=~7~\TeV{} with the ATLAS detector at the LHC during 2011.
  Events dominated by $W\gamma$ and $Z\gamma$ production with leptonic decays of the
  $W$ and $Z$ bosons are selected, and their production cross sections and kinematic
  properties are measured in several ranges of the photon transverse
  energy. The results are compared to Standard Model predictions and
  are used to determine limits on anomalous $WW\gamma$ and
  $ZZ\gamma$/$Z\gamma\gamma$ couplings.
}  

\PreprintJournalName{Physics Letters B}  

\usepackage{amssymb}

\usepackage{atlasphysics}
\usepackage{subfigure}
\usepackage{graphicx}

\usepackage{longtable}
\usepackage{multirow}
\usepackage{textcomp}
\usepackage{units}
\usepackage{marvosym}
\usepackage{hyperref}

\usepackage{graphicx}
\usepackage{dcolumn}
\usepackage{bm}
\usepackage[mathlines]{lineno}

\begin{document}
\title{Measurement of $W\gamma$ and $Z\gamma$ production cross sections in $pp$ collisions at $\sqrt{s}=$ 7 TeV and limits on anomalous triple gauge couplings with the ATLAS detector}

\collaboration{ATLAS Collaboration}
\date{May 11, 2012}

\begin{abstract}

  This Letter presents measurements of $l^{\pm}\nu\gamma$ and
  $l^{+}$$l^{-}$$\gamma$ ($l=e$, $\mu$) production in 1.02 fb$^{-1}$ of $pp$ collision data
  recorded at $\sqrt{s}$~=~7~\TeV{} with the ATLAS detector at the LHC during 2011.
  Events dominated by $W\gamma$ and $Z\gamma$ production with leptonic decays of the
  $W$ and $Z$ bosons are selected, and their production cross sections and kinematic
  properties are measured in several ranges of the photon transverse
  energy. The results are compared to Standard Model predictions and
  are used to determine limits on anomalous $WW\gamma$ and
  $ZZ\gamma$/$Z\gamma\gamma$ couplings.
\end{abstract}

\maketitle


\section{Introduction}
\label{sec:intro}

The Standard Model (SM) predicts self-couplings of the $W$ boson, the
$Z$ boson and the photon through the non-Abelian $SU(2)_L\times
U(1)_Y$ gauge group of the electroweak sector.  Experimental tests of
these predictions have been made in $p\bar{p}$ and $pp$ collider
experiments through the $s$-channel production of one of the gauge
bosons and its subsequent coupling to a final state boson pair such as
$WW$, $WZ$, and $W\gamma$ ($s$-channel production of $ZZ$ and
$Z\gamma$ are forbidden in the SM). The production cross sections are
sensitive to the couplings at the triple gauge-boson (TGC) vertices and
therefore provide direct tests of SM predictions.  Deviations of the 
TGC from the SM expectation could occur from a composite structure of
the $W$ and $Z$ bosons, or from the presence of new bosons that decay
to SM vector boson pairs. Previous measurements of $W\gamma$ and
$Z\gamma$ production have been made at the Tevatron by the
CDF~\cite{CDFpaper} and D0~\cite{D0paper,D0paper2}
collaborations, and at the CERN Large Hadron Collider (LHC) by the
ATLAS ~\cite{ATLASpaper} and CMS~\cite{CMSpaper} collaborations.

In this Letter we report measurements of the production of $W\gamma$
and $Z\gamma$ boson pairs from $pp$ collisions provided by the LHC, 
at a centre-of-mass energy of 7 TeV.  The analysis presented
here uses a data sample corresponding to an integrated luminosity of 1.02
fb$^{-1}$ collected by the ATLAS experiment in 2011. Events
triggered by high transverse energy ($E_{\mathrm{T}}$) electrons and
high transverse momentum ($p_{\mathrm{T}}$) muons are used to select
$pp \to l^{\pm}  \nu  \gamma + X$ and $pp \to l^+  l^-  \gamma +
X$ production.
Several processes contribute to these final states, including final state radiation (FSR) of photons from charged leptons in inclusive $W$ or $Z$ production, radiation of photons from initial or final state quarks in $W$ or $Z$ production, and radiation of
photons directly from $W$ bosons through the $WW\gamma$ vertex.

The production processes are categorized according to the photon transverse energy. The event sample
with low $E_{\mathrm{T}}^\gamma$ photons includes a large contribution from $W$/$Z$ boson decays with final state radiation.
For a better comparison to SM predictions the events are analyzed both inclusively, with no requirements on the recoil system, and exclusively, requiring that there is no hard jet. The inclusive $V\gamma$ ($V=W$ or $Z$) event sample includes significant contributions of photons from final state parton fragmentation, whereas for exclusive $V\gamma$ events, the photons originate primarily as radiation from initial state quarks in $W$ and $Z$ production, or from the $WW\gamma$ vertex in $W\gamma$ events.
The measurements of exclusive $V\gamma$ events with high $E_{\mathrm{T}}^\gamma$ photons are used to extract limits on anomalous triple gauge-boson couplings (aTGCs). The observed limits are compared with the corresponding measurements at the Tevatron~\cite{CDFpaper,D0paper,D0paper2} and LEP~\cite{LEP}, as well as the measurements from CMS~\cite{CMSpaper}.


\section{The ATLAS Detector and the Data Sample}
\label{sec:atlasdet}
The ATLAS detector~\cite{DetectorPaper:2008} is composed of
an inner tracking system (ID) surrounded by a thin
superconducting solenoid providing a 2~T axial magnetic field,
electromagnetic (EM) and hadronic calorimeters, and a muon spectrometer (MS).  
The ID consists of three subsystems: the pixel and silicon
microstrip (SCT) detectors cover the pseudorapidity range $|\eta|<2.5$ 
\footnote{ATLAS uses a right-handed coordinate system with its origin at the nominal interaction point (IP) in the centre of the detector and the $z$-axis along the beam pipe. The $x$-axis points from the IP to the centre of the LHC ring, and the $y$ axis points upward. Cylindrical coordinates $(r,\phi)$ are used in the transverse plane, $\phi$ being the azimuthal angle around the beam pipe. The pseudorapidity is defined in terms of the polar angle $\theta$ as $\eta=-\ln\tan(\theta/2)$. The distance $\Delta R$ in the $\eta-\phi$ space is defined as $\Delta R = \sqrt{({\Delta \eta})^2 + ({\Delta \phi})^2}$}, while the Transition Radiation Tracker (TRT) has an acceptance range of $|\eta|<2.0$.
The calorimeter system covers the range $|\eta|<4.9$ and is composed of
sampling calorimeters with either liquid argon (LAr) or scintillating tiles
as the active media. In the region $|\eta|<2.5$ the EM LAr calorimeter
is finely segmented and plays an important role in electron and photon identification.
The MS is based on three large superconducting toroids arranged with an eight-fold azimuthal coil symmetry around the calorimeters, and a system of three stations of chambers for the trigger and precise measurements of muon tracks.
Data were collected during the first half of 2011 from 
$pp$ collisions.
Events were selected by triggers requiring at least one identified
electron with $E_{\mathrm{T}}>20$~GeV{}
or a muon with $p_{\mathrm{T}}>18$~GeV{}.
The total integrated luminosity used for this measurement is 1.02 fb$^{-1}$ with an
uncertainty of 3.7$\%$~\cite{ATLASLumi1,ATLASLumi2}.


\section{Simulation of $W\gamma$ and $Z\gamma$ events and backgrounds}
\label{sec:simulation}
 Monte Carlo (MC) event samples, including a full simulation~\cite{ATLASsim}
of the ATLAS detector within {\sc geant4}~\cite{geant4},
are used to compare the data to the SM signal and background expectations.
All MC samples are simulated with in-time pile-up (an average
of four $pp$ interactions within a single bunch crossing) and
out-of-time pile-up (signals from neighbouring bunch crossings). The
simulated events are weighted such that the distribution of multiple
collisions per bunch crossing matches what is observed in the data
for the period used in this analysis.

The production $pp \to l^{\pm} \nu  \gamma +X$ is modelled with the
{\sc alpgen} generator~\cite{alpgen} interfaced to
{\sc herwig}~\cite{herwig} for parton shower and fragmentation processes,
and to {\sc jimmy}~\cite{jimmy} for underlying event simulation. The
{\sc sherpa} generator~\cite{sherpa} is used to simulate $pp \to l^+ 
l^- \gamma + X$ events.  The {\sc cteq6l1}~\cite{cteq6l1} and
{\sc cteq6.6m}~\cite{cteq66m} parton distribution functions (PDF) are used for
samples generated with {\sc alpgen} and {\sc sherpa}, respectively.
The FSR photons from charged leptons is handled by
{\sc photos}~\cite{photos} for the {\sc alpgen} sample,
and by the {\sc sherpa} generator for the {\sc sherpa} sample.
All the signal production processes, including the photon
fragmentation, are simulated by these two generators.  The {\sc alpgen}
sample is generated with leading-order (LO) matrix elements for final
states with up to five partons, whereas the {\sc sherpa} sample is
generated with LO matrix elements for final states with up
to three partons.  
%
The $Z \rightarrow ll$ and $W \rightarrow \tau \nu$ backgrounds are
modelled with {\sc pythia}~\cite{pythia}.
The radiation of photons from charged leptons is treated in {\sc pythia}
using {\sc photos}.
{\sc tauola}~\cite{tau} is used for $\tau$ lepton decays.
The {\sc powheg}~\cite{powheg} generator is used to simulate $t\bar{t}$ production,
interfaced to {\sc pythia} for parton showering. 
The $WW$ and single-top quark productions are modelled by {\sc
  mc@nlo}~\cite{MCatNLO1,MCatNLO2}, interfaced to {\sc herwig}
for parton showering and fragmentation. The next-to-leading-order (NLO) 
cross-section predictions are used to normalize the simulated
background events. Other backgrounds are derived from data as
described in Section~\ref{sec:background}.


\section{Reconstruction and selection of $W\gamma$ and $Z\gamma$ candidates}
\label{sec:Event_Selection}

The $W$ and $Z$ bosons are selected through their decays into $e\nu$, $\mu\nu$ and $e^{+}e^{-}$, $\mu^{+}\mu^{-}$ respectively.
The $W\gamma$ final state consists of an isolated electron or muon,
large missing transverse momentum due to the undetected neutrino, and an isolated photon.
The $Z\gamma$ final state contains one $e^{+}e^{-}$ or $\mu^{+}\mu^{-}$ pair and an isolated photon.
Collision events are selected by requiring at least one reconstructed vertex with at 
least three charged particle tracks.
If more than one vertex satisfies the vertex selection requirement,
the vertex with the highest sum of the $p_{\mathrm{T}}^{2}$ of the associated tracks is chosen.

An electron candidate is obtained from an energy cluster in the EM calorimeter associated with a reconstructed charged particle in the ID.
The electron's $E_{\mathrm{T}}$ must be greater than 25 GeV{}.
To avoid the transition regions between the calorimeters, the electron
cluster must satisfy $|\eta|<1.37$ or $1.52<|\eta|<2.47$.
The selection of $W(\rightarrow e \nu)\gamma$ events requires one electron passing tight identification cuts~\cite{atlas_electron}.
Two oppositely charged electrons passing medium identification cuts~\cite{atlas_electron} are required in the $Z(\rightarrow e^+ e^-)\gamma$ selection.
To reduce the background due to a jet misidentified as an electron
in the $W\gamma$ analysis,
a calorimeter-based isolation requirement $E_{\mathrm{T}}^{\mathrm{iso}}<6$ GeV{} is applied
to the electron candidate.
$E_{\mathrm{T}}^{\mathrm{iso}}$ is the total transverse energy recorded in the
calorimeters within a cone of radius $\Delta R=0.3$ around the electron direction
(excluding the energy from the electron cluster).
$E_{\mathrm{T}}^{\mathrm{iso}}$ is corrected for leakage of the electron energy outside the electron cluster and for contributions from the underlying event and pile-up~\cite{photonpaper}.


Muon candidates are identified by associating complete tracks or track segments in the MS to tracks in the ID~\cite{WZpaper}.
Each selected muon candidate is a combined track originating from the primary vertex
with transverse momentum $p_{\mathrm{T}}>25$ GeV and $|\eta|<2.4$.
It is required to be isolated by imposing $R^{\mathrm{iso}}(\mu)<0.1$,
where $R^{\mathrm{iso}}(\mu)$ is the sum of the track $p_{\mathrm{T}}$
in a $\Delta R =0.2$ cone around the muon direction divided by the muon $p_{\mathrm{T}}$.
For the $W(\rightarrow \mu \nu)\gamma$ measurement at least one muon candidate
is required in the event, whereas for the $Z(\rightarrow \mu^+\mu^-)\gamma$ measurement, the
selected events must have exactly two oppositely charged muon candidates.

Photon candidates use clustered energy deposits in the EM calorimeter in the
range $|\eta|<2.37$ (excluding the calorimeter transition region $1.37<|\eta|<1.52$) with $E_{\mathrm{T}}>$ 15 \GeV{}.
Requirements on the shower shape~\cite{photonpaper} are
applied to suppress the background from multiple showers produced in
meson (e.g. $\pi^{0}$, $\eta$) decays. To further reduce this
background, a photon isolation requirement $E_{\mathrm{T}}^{\mathrm{iso}}<6$ GeV is applied.  The definition of
photon isolation is similar to the electron isolation described above.

The reconstruction of the missing transverse momentum ($E_{\mathrm{T}}^{\mathrm{miss}}$)~\cite{METpaper} is based on the energy deposits in calorimeter cells inside three-dimensional clusters. Corrections for the calorimeter response to hadrons, dead material, out-of-cluster energy, as well as muon momentum are applied.
A selection requirement of $E_{\mathrm{T}}^{\mathrm{miss}}>25$ GeV is applied in the
$W\gamma$ analysis.
Jets are reconstructed from calorimeter clusters using the anti-$k_t$ jet clustering
algorithm~\cite{antikt} with radius parameter $R=0.4$.
The selected jets are required to have $p_{\mathrm{T}}>30$ GeV with $|\eta| < 4.4$,
and to be well separated from the lepton and photon candidates
($\Delta R(e/\mu/\gamma,$ jet)$>0.6$).
In the exclusive $W\gamma$ and $Z\gamma$ analyses, events with one or more jets are vetoed.

For each selected $W\gamma$ candidate event, in addition to the presence of one high
$p_{\mathrm{T}}$ lepton, one high $E_{\mathrm{T}}$ isolated photon and large $E_{\mathrm{T}}^{\mathrm{miss}}$,
the transverse mass of the lepton-$E_{\mathrm{T}}^{\mathrm{miss}}$ system
is required to be $m_{\mathrm{T}}(l,\nu)= \sqrt{2p_{\mathrm{T}}(l) \cdot E_{\mathrm{T}}^{\mathrm{miss}} \cdot (1-\cos{\Delta{\phi}})} > 40$ \GeV{}, where $\Delta{\phi}$ is the azimuthal separation between the directions of the lepton
and the missing transverse momentum vector. A $Z$-veto requirement is applied in the electron channel of the $W\gamma$ analysis by asking that the electron-photon invariant mass ($m_{e\gamma}$) is not within 10~\GeV{} of the $Z$-boson mass.

For $Z\gamma$ candidates, the invariant mass of the two oppositely charged
leptons is required to be greater than 40~\GeV{}.
In both $W\gamma$ and $Z\gamma$ analyses, a requirement $\Delta R(l,\gamma)>$ 0.7 is applied to suppress the contributions from FSR photons in $W$ and $Z$ boson decays.


\section{Signal efficiencies}   
\label{sec:eff}
The efficiencies of the lepton selections,
and the lepton triggers, are first estimated from the $W/Z+\gamma$ signal MC events
and then corrected with scale factors derived using  high purity lepton data samples from $W$ and $Z$ boson decays to account for small discrepancies between the data and the MC simulation~\cite{atlas_electron,WZpaper,ATLAS_Wjet,photonpaper}.

The average efficiency for the tight electron selection in $W\gamma$ events is $(74.9 \pm 1.2)\%$.
For the medium quality electron selection in $Z\gamma$ events, the efficiency is $(96.4 \pm 1.4)\%$
and $(91.0 \pm 1.6)\%$ for the leading and sub-leading electron, respectively.
The electron-isolation efficiency is $>99\% \pm 1\%$.
The uncertainties reported throughout this Letter, unless stated otherwise, reflect the combined statistical and systematic uncertainties.
The efficiency of the electron trigger, which is used to select the data sample
for the electron decay channels, is found to be $>99.5\%$ for both tight and medium electron candidates.

The muon-identification efficiency for the
$W\gamma$ and $Z\gamma$ analyses is estimated to be $(90 \pm 1)\%$.
The muon-isolation efficiency is $>99\%$ with negligible uncertainty.
The efficiency of the muon trigger to select the $W\gamma$ and $Z\gamma$ events
is $(83 \pm 1)\%$ and $(97 \pm 1)\%$, respectively. 

The photon identification efficiency is determined from $W\gamma$ and $Z\gamma$
MC samples where the shower shape distributions are corrected
to account for the observed small discrepencies between data and simulation.
The photon identification efficiency increases with the
photon $E_{\mathrm{T}}$, and is estimated to be $68\%$, $88\%$ and
$90\%$ for photons with $E_{\mathrm{T}}>15$, 60 and 100 GeV{},
respectively. 
 The main sources of systematic uncertainty come from
the imperfect knowledge of the material in front of the calorimeter, 
the background contamination in the samples used to determine the corrections to the
shower shape variables, and pile-up effects~\cite{photonpaper}.  The systematic uncertainty in the identification
efficiency due to the uncertainty in the photon contributions from
quark/gluon fragmentation is also considered.
The overall relative uncertainty in the photon identification efficiency is 11\% for $E_{\mathrm{T}}>15$ GeV{}, decreasing to 4.5\% for $E_{\mathrm{T}}>$ 60 or 100 GeV{}.
The photon isolation efficiency is estimated using $W\gamma$ and $Z\gamma$ signal MC events
and cross-checked with data using electrons from $Z \rightarrow e^{+}e^{-}$ decays~\cite{atlas_electron}.
The estimated efficiency varies from (98 $\pm$ 1.5)\% for $E_{\mathrm{T}}>15$ GeV{} to (91 $\pm$ 2.5)\% for $E_{\mathrm{T}}>100$ GeV{}.

\section{Background Determination and Signal Yield}
\label{sec:background}

The dominant source of background in this analysis comes from
$V+$jets ($V$=$W$ or $Z$) events where photons from the decays of mesons
produced in jet fragmentation (mainly $\pi^0 \to \gamma \gamma$)
pass the photon selection criteria.
Since the fragmentation functions of quarks and gluons into hadrons are
poorly constrained by experiments,
these processes may not be well modelled by the MC simulation. 
Therefore the $V+$jets backgrounds are derived from data.

For the $W\gamma$ analysis, another important source of background which is not well modelled by MC simulations is the $\gamma+$jets process. These background events can be misidentified as
$W\gamma$ events when there are leptons from heavy quark decays
(or the hadrons inside jets are misidentified as leptons) and large apparent
$E_{\mathrm{T}}^{\mathrm{miss}}$ is created by the mis-measurement of the jet energies.

The background contributions from $W+$jets and $\gamma+$jets events in
the $W\gamma$ analysis, or from $Z+$jets events in the $Z\gamma$
analysis, are estimated from data. 

The $Z \rightarrow l^+l^-$ process is also one of the dominant backgrounds in the
$W\gamma$ analysis. Its contribution is estimated from MC simulation,
since this process is well understood and modelled.  
Other backgrounds such as those from $t\bar{t}$ decay for the $Z\gamma$ analysis, and those from electroweak (EW) processes ($W
\rightarrow\tau \nu$, $WW$), single top and $t\bar{t}$ for the $W\gamma$ analysis, are less important and are estimated from MC simulation.
 These processes, together
with the $Z \rightarrow l^+l^-$ background, are referred to collectively as ``EW+$t\bar t$ background''.


The misidentified photons (leptons) in $V+$jets ($\gamma+$jets) events are more likely to fail the photon (lepton) isolation criteria.
A ``pass-to-fail'' ratio $f_\gamma$ ($f_l$) is defined as the ratio of photon (lepton) candidates passing the photon (lepton) isolation criteria to the number of candidates failing the isolation requirement. 
The ratio $f_\gamma$ is measured in $W~\rightarrow l\nu$ ($Z~\rightarrow l^+ l^-$) events with one ``low quality'' photon candidate. A ``low quality'' photon candidate is defined as one that fails the photon shower-shape selection criteria, but passes a background-enriching subset of these criteria. The ratio $f_e$ is measured in a control sample, which requires the events to pass all the $W+\gamma$ selection criteria, except the $E_{\mathrm{T}}^{\mathrm{miss}}$ requirement. The control sample for $f_\mu$ measurement is defined in a way similar to that used for $f_e$, except that in addition the muon track is required to have a large impact parameter in order to enhance the heavy flavor component.
%
  The estimated contribution of $V+$jets is obtained by multiplying the measured $f_\gamma$ by the number of events passing all $V+\gamma$ selections, except the photon isolation requirement. Similarly the $\gamma+$jets background is estimated using the measured $f_l$.


\begin{figure*}
  \centering
  \subfigure[]{\includegraphics[width=0.99\columnwidth]{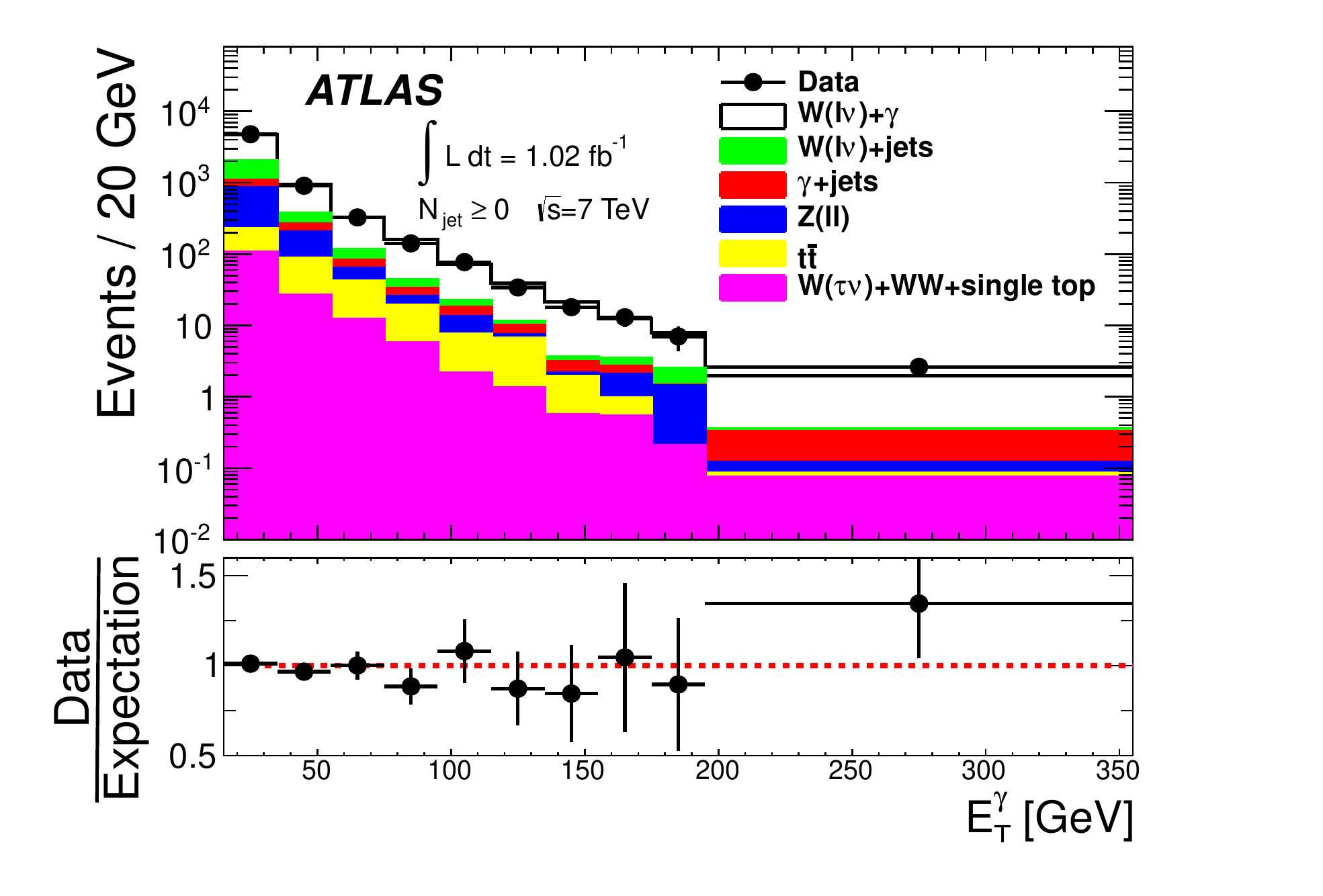}}
  \subfigure[]{\includegraphics[width=0.99\columnwidth]{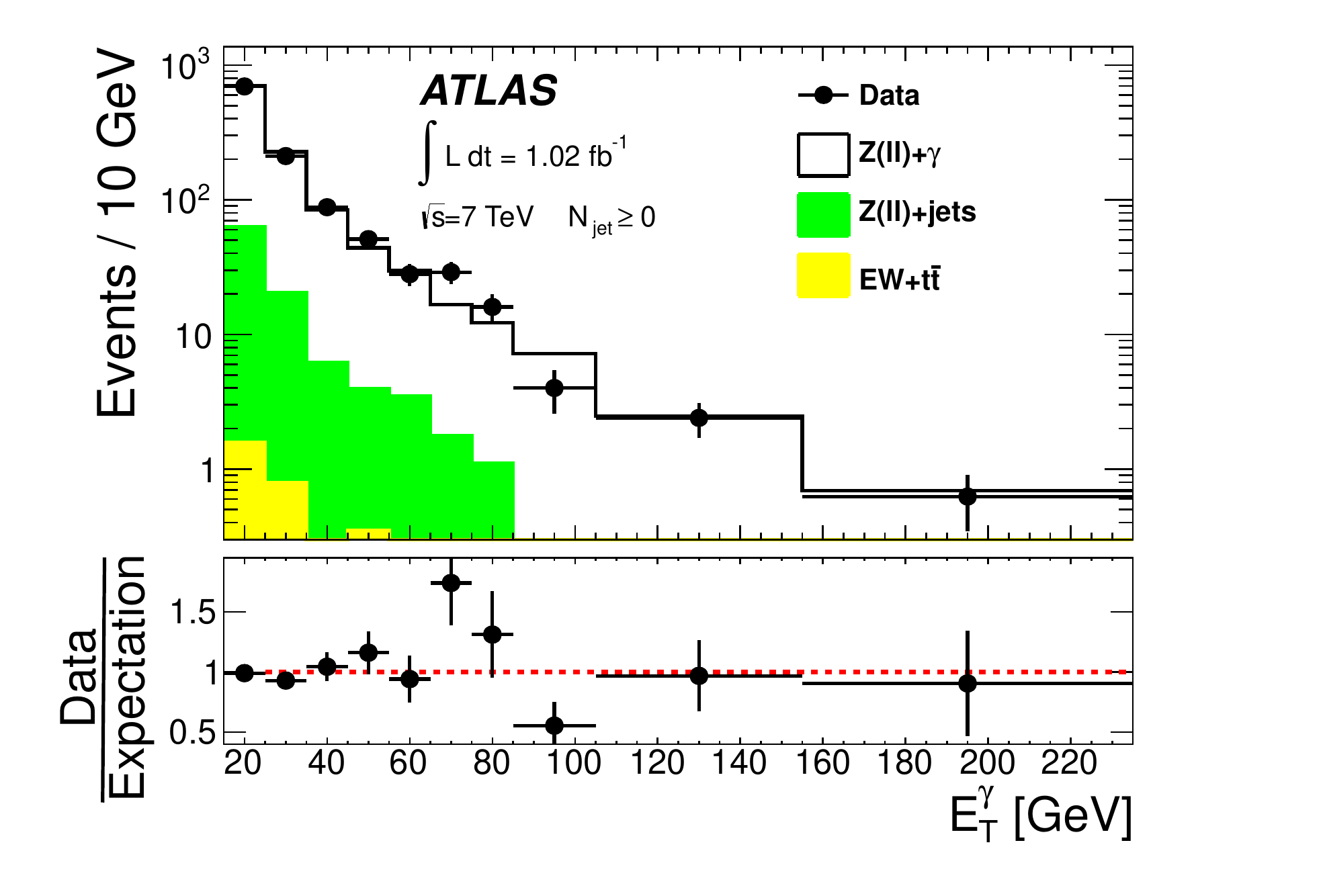}}



  \caption{Distributions of the photon transverse energy for the combined
electron and muon decay channels in (a) $W\gamma$ candidate events and (b) $Z\gamma$ candidate events, with no requirements on the recoil system. The selection criteria are defined in Section~\ref{sec:Event_Selection}. The distributions for the expected signals are taken from the MC simulation and normalised to the extracted number of signal events shown in Table~\ref{tab:wgbg} and Table~\ref{tab:zgbgsummary}. The ratio between the number of candidates observed in the data and the number of expected candidates from the signal MC simulation and from the background processes is also shown.}

  \label{fig:Wg_kin_elec1}
\end{figure*}

\begin{figure*}
  \centering
  \subfigure[]{\includegraphics[width=0.99\columnwidth]{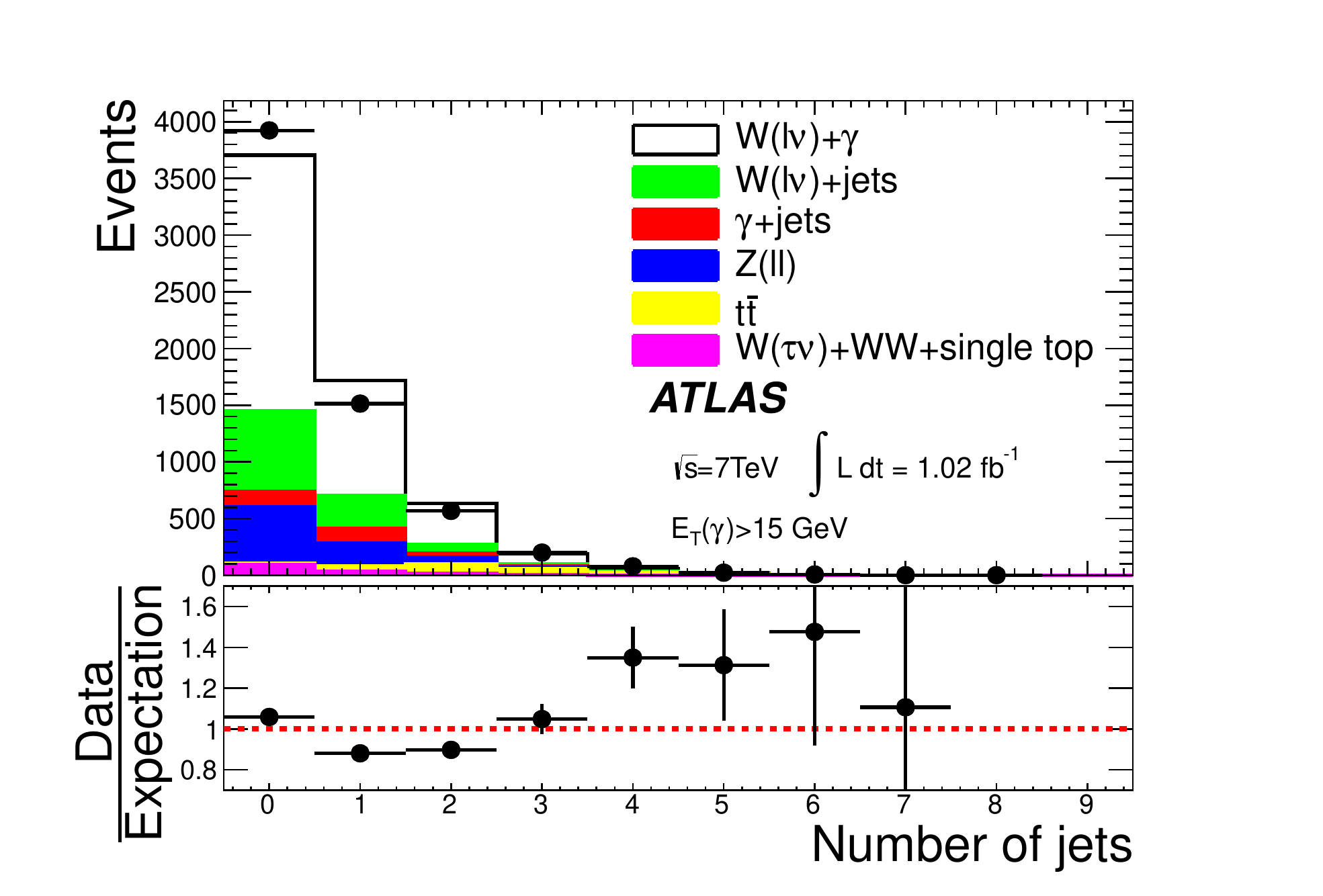}}
  \subfigure[]{\includegraphics[width=0.99\columnwidth]{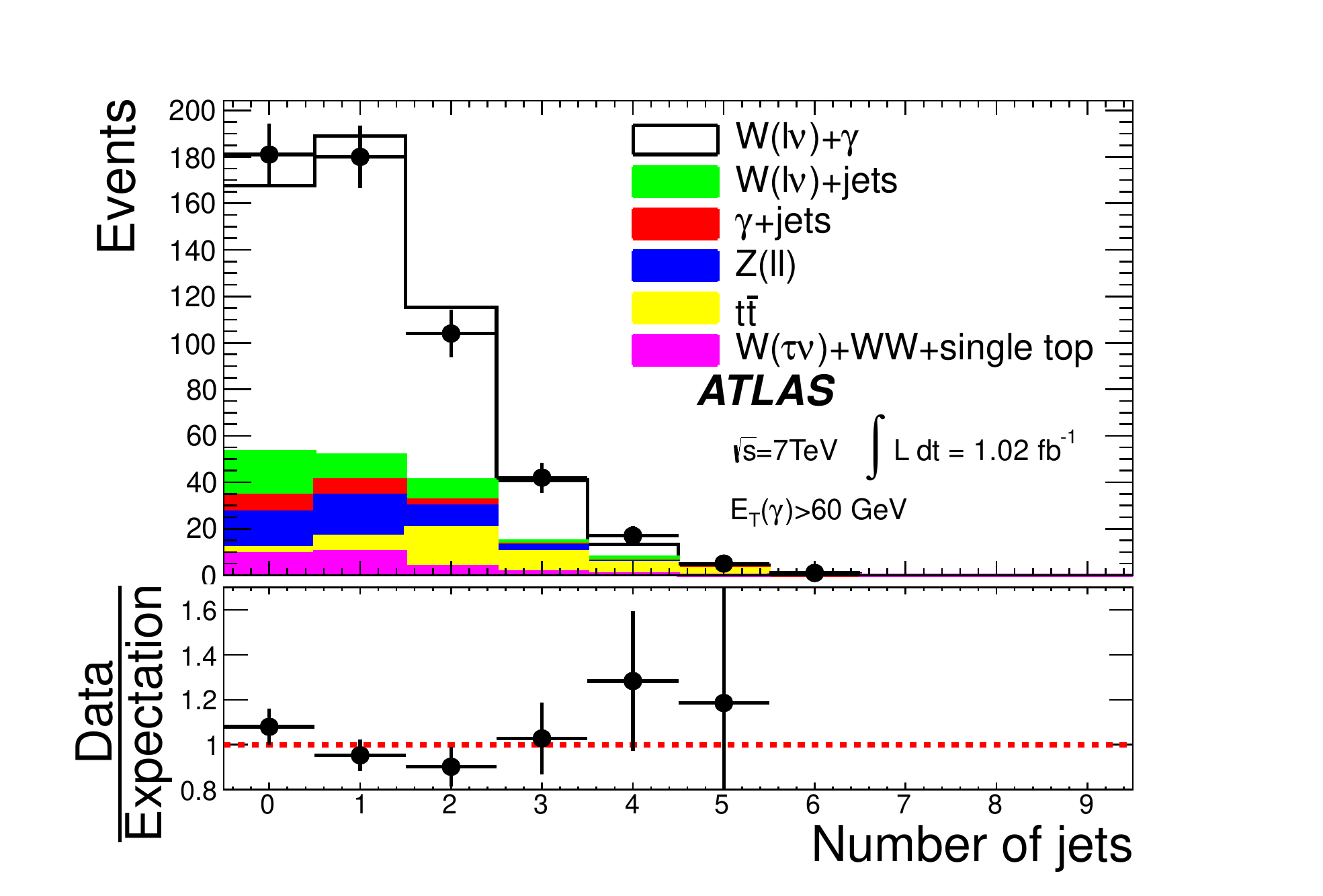}}
  \subfigure[]{\includegraphics[width=0.99\columnwidth]{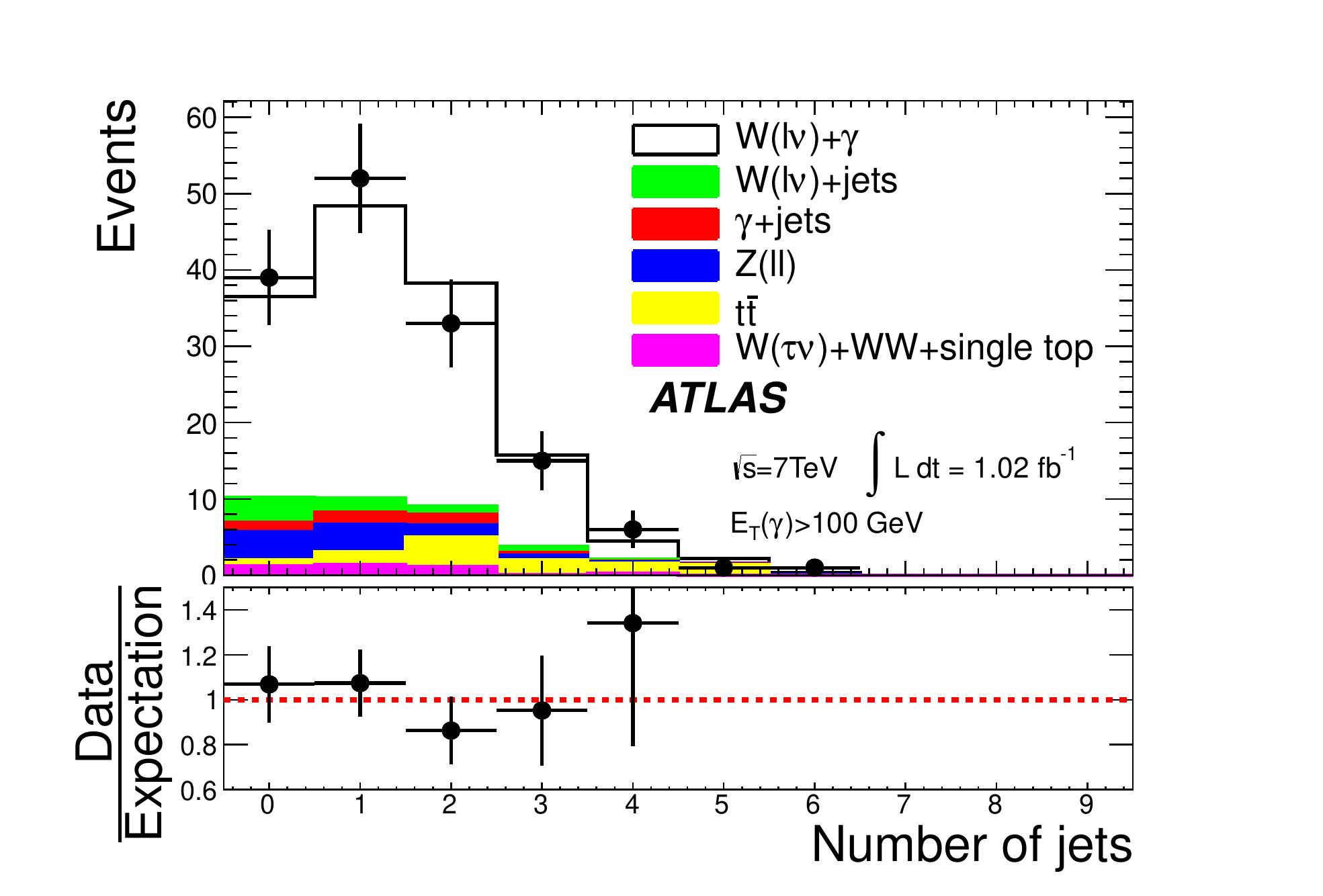}}
  \subfigure[]{\includegraphics[width=0.99\columnwidth]{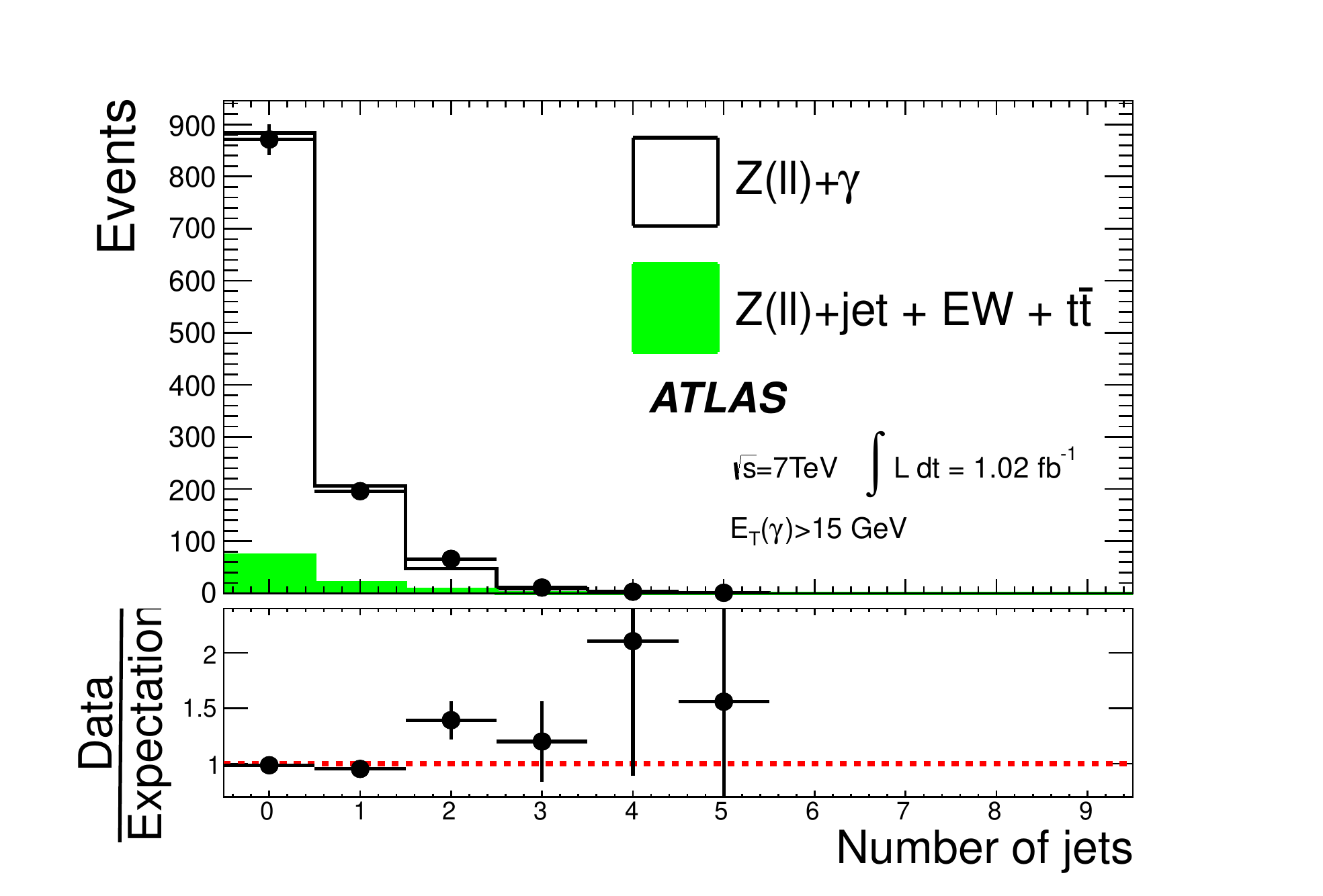}}
  \subfigure[]{\includegraphics[width=0.99\columnwidth]{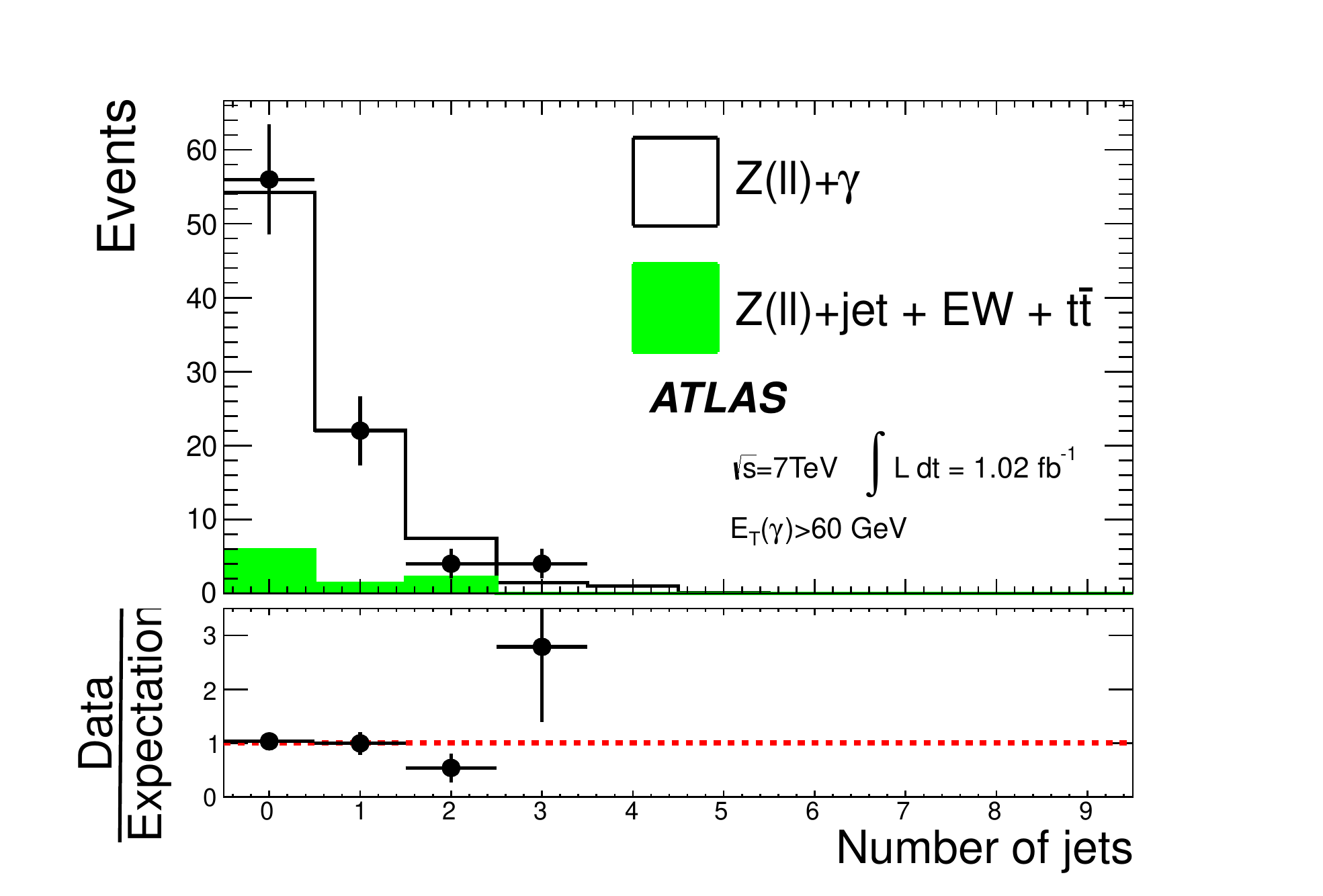}}

  \caption{Distributions of the jet multiplicity for the combined
electron and muon decay channels in (a) $W\gamma$ candidate events with $E_{\mathrm{T}}^\gamma>15$ \GeV{}, (b) $W\gamma$ candidate events with $E_{\mathrm{T}}^\gamma>60$ \GeV{}, (c) $W\gamma$ candidate events with $E_{\mathrm{T}}^\gamma>100$ \GeV{}, (d) $Z\gamma$ candidate events with $E_{\mathrm{T}}^\gamma>15$ \GeV{}, and (e) $Z\gamma$ candidate events with $E_{\mathrm{T}}^\gamma>60$ \GeV{}. The selection criteria are defined in Section~\ref{sec:Event_Selection}. Distributions for expected signal contribution are taken from signal MC simulation and normalized to the extracted number of signal events as shown in Table~\ref{tab:wgbg} and Table~\ref{tab:zgbgsummary}. The ratio between the number of candidates observed in the data and the number of expected candidates from the signal MC simulation and from the background processes is also shown.}
  \label{fig:Wg_kin_elec3}
\end{figure*}

The accuracy of the $W/Z/\gamma+$jets background determination has been assessed in detail. The ratios $f_\gamma$ and $f_l$, which are measured in background-enriched samples, may be biased due to the different composition of these samples and the signal sample. To estimate the uncertainty in $f_\gamma$ from this source, two sets of alternative selections, with tighter and looser background selection requirements, are used to obtain alternative control samples. $f_e$ is also measured in an alternative control sample selected by requiring that events pass all $W$+$\gamma$ selection criteria, except that the electron fails the tight identification criteria but passes the low quality criteria.  
To determine the systematic uncertainty on $f_\mu$, the $E_{\mathrm{T}}^{\mathrm{miss}}$ and impact parameter requirements for the muon track are varied to obtain alternative control samples.
The $W/Z/\gamma+$jets background estimates from the alternative control samples are consistent with those obtained from the nominal samples, and the differences are assigned as systematic uncertainties. The changes in the background estimates from varying the photon or lepton isolation requirements are also assigned as systematic uncertainties.

Extrapolation methods are used to cross-check the $W/Z/\gamma+$jets background estimates in the high $E_{\mathrm{T}}^{\gamma}$ region, where few events are available. The extrapolation method scales the well-measured background level in the low $E_{\mathrm{T}}^{\gamma}$ region to the high $E_{\mathrm{T}}^{\gamma}$ region using the $E_{\mathrm{T}}^{\gamma}$ distribution shape obtained from control samples.
The differences between results obtained from the nominal and extrapolation methods are used as additional uncertainties.

The uncertainties on the "tt+EW" background include the theoretical uncertainty on the NLO cross section (between 6\%--7\% depending on the process), the luminosity uncertainty (3.7\%)~\cite{ATLASLumi1,ATLASLumi2} and the experimental systematic uncertainty. The latter is dominated by the uncertainties on the uncertainty on the jet energy scale (5\%) and the uncertainty on the EM shower shape modelling in the MC simulation (4\%--11\%).

A summary of background contributions and signal yields in the $W\gamma$ and $Z\gamma$ analyses is given in Table~\ref{tab:wgbg} and Table~\ref{tab:zgbgsummary} respectively. The photon transverse energy and jet multiplicity distributions from the selected $W\gamma$ and $Z\gamma$ events are shown in Figure~\ref{fig:Wg_kin_elec1} and Figure~\ref{fig:Wg_kin_elec3}, respectively. The data are compared to the sum of the backgrounds and the SM signal predictions. 
The distributions for the expected $W\gamma$ and $Z\gamma$ signal are taken from signal MC simulation and normalized to the extracted number of signal events shown in Table~\ref{tab:wgbg} ($N^{\mathrm{sig}}_{W\gamma}$) and Table~\ref{tab:zgbgsummary} ($N^{\mathrm{sig}}_{Z\gamma}$).

\begin{table}[htbp]


 \begin{tabular}{ccccccccc}
       \hline
       \hline
      \textbf{} & $pp \rightarrow e\nu \gamma$  & $pp \rightarrow \mu\nu \gamma$ & $pp \rightarrow e\nu \gamma$  & $pp \rightarrow \mu\nu \gamma$ \\
       \hline
       \hline
        Region  &\multicolumn{2}{c}{ $E_{\mathrm{T}}^\gamma> 15$ \GeV{}}&\multicolumn{2}{c}{$E_{\mathrm{T}}^\gamma> 15$ \GeV{}} \\
           &\multicolumn{2}{c}{$N_{\mathrm{jet}}\geq0$ }&\multicolumn{2}{c}{$N_{\mathrm{jet}}=0$} \\

       \hline
       $N^{\mathrm{obs}}_{W\gamma}$ &  2649            &      3621        &        1666       & 2238    \\ 
       \hline
        $W$+jets                & $439  \pm 108$ & $685 \pm 162$ & $242\pm 68$  & $473 \pm 128$   \\  
        $\gamma$+jets           & $255  \pm 58 $ & $67 \pm 16 $   & $119  \pm 34$  & $28.9\pm 7.4$    \\  
        EW                & $405\pm 53$  & $519 \pm 67$  & $229 \pm 30$  & $366 \pm 48$    \\  
        $t\bar t$              &  $85\pm 11 $   & $152 \pm 20$  & $1.6 \pm0.4$      & $8.1 \pm 1.3$    \\  
       \hline
       $N^{\mathrm{sig}}_{W\gamma}$        & $1465 \pm 139$  & $2198  \pm 183$  & $1074 \pm 91$     & $1362 \pm 145$ \\

       \hline
       \hline
        Region  &\multicolumn{2}{c}{ $E_{\mathrm{T}}^\gamma> 60$ \GeV{}}&\multicolumn{2}{c}{$E_{\mathrm{T}}^\gamma> 60$ \GeV{}} \\
           &\multicolumn{2}{c}{$N_{\mathrm{jet}}\geq0$ }&\multicolumn{2}{c}{$N_{\mathrm{jet}}=0$} \\

       \hline
      $N^{\mathrm{obs}}_{W\gamma}$    &  216             &        307       &   76              &  104    \\ 
       \hline
        $W$+jets                & $14.2  \pm 6.9$   & $27.1\pm 10.1$  &  $6.4  \pm 3.5$   & $12.9  \pm 5.9$  \\  
        $\gamma$+jets           & $10.8 \pm 6.6$   & $7.1   \pm 5.1$   &  $5.5 \pm 4.2$   & $1.7    ^{+4.5}_{-0.7}$   \\  
        EW                     & $32.0 \pm 3.6$   & $29.9 \pm 3.6$  &  $9.2 \pm 1.8$    & $12.6 \pm 2.0$   \\  
        $t\bar t$              & $13.1 \pm 1.4$   & $29.5 \pm 3.2$   & $0.3 \pm 0.2$    & $2.4 \pm 0.6$   \\  
       \hline
       $N^{\mathrm{sig}}_{W\gamma}$        &  $146  \pm 16$ & $214  \pm 19$ &  $54.6  \pm 9.4$  & $74.4 \pm 11.4$ \\

\hline
       \hline
        Region  &\multicolumn{2}{c}{ $E_{\mathrm{T}}^\gamma> 100$ \GeV{}}&\multicolumn{2}{c}{$E_{\mathrm{T}}^\gamma> 100$ \GeV{}} \\
           &\multicolumn{2}{c}{$N_{\mathrm{jet}}\geq0$ }&\multicolumn{2}{c}{$N_{\mathrm{jet}}=0$} \\
       \hline
     $N^{\mathrm{obs}}_{W\gamma}$    &    61            &      85          &   21              &  18   \\ 
       \hline
        $W$+jets                & $4.5 \pm 2.8$    & $2.8 \pm 2.1$  & $2.9    \pm 2.2$   & $0.4 ^{+0.7}_{-0.4}$   \\  
        $\gamma$+jets           &  $2.4\pm 2.4$   & $2.4 ^{+2.7}_{-2.4}$    &  $1.0  ^{+2.3}_{-1.0}$   & $0.2  ^{+0.7}_{-0.2}$    \\  
        EW                     &   $5.8\pm 1.1$   & $8.0 \pm 1.8$    & $2.5 \pm 0.8$     & $4.0 \pm 1.1$    \\  
        $t\bar t$              &   $3.4 \pm 0.6$  & $7.6 \pm 0.9$    & $0.2 \pm 0.1$     & $0.6 \pm 0.3$    \\  
       \hline
       $N^{\mathrm{sig}}_{W\gamma}$        & $44.9 \pm 7.7$  & $64.2  \pm 8.9 $ &      $14.4  \pm 5.0$    & $12.8  \pm 3.8$    \\  
\hline
       \hline
\end{tabular}

\caption{
Expected numbers of background events, observed numbers of signal events ($N^{sig}_{W\gamma}$) and total numbers of events passing the selection
    requirements in the data ($N^{obs}_{W\gamma}$) for the $pp \rightarrow e\nu \gamma$ channel and the $pp \rightarrow \mu\nu \gamma$ channel in different $E_{\mathrm{T}}^\gamma$ and jet multiplicity regions. The combined statistical and systematic uncertainties are shown. The uncertainty on the background prediction is dominated by systematic uncertainties in all regions. The contribution from the EW background is dominated by the $Z \rightarrow e^+e^-(\mu^+ \mu^-)$ process.}
  \label{tab:wgbg} 
\end{table}

\begin{table}[htbp]

 \begin{tabular}{ccccccccc}
       \hline
       \hline
      \textbf{} & $e^{+}e^{-}\gamma$  & $ \mu^{+}\mu^{-}\gamma$ &
                                    $ e^{+}e^{-}\gamma$  & $\mu^{+}\mu^{-}\gamma$ \\
       \hline
        Region  &\multicolumn{2}{c}{ $E_{\mathrm{T}}^\gamma> 15$ \GeV{}}&\multicolumn{2}{c}{$E_{\mathrm{T}}^\gamma> 15$ \GeV{}} \\
           &\multicolumn{2}{c}{$N_{\mathrm{jet}}\geq0$ }&\multicolumn{2}{c}{$N_{\mathrm{jet}}=0$} \\

       \hline
       $N^{\mathrm{obs}}_{Z\gamma}$  & 514    &  634              & 376    &  495    \\ 
       \hline
        $N_{Z\gamma}^{\mathrm{BG}}$             & $43.7  \pm 16.5 $ &  $56.8  \pm 16.2$  & $ 29.3  \pm 11.0  $  &  $39.3 \pm 15.8$    \\ 
       \hline
       $N^{\mathrm{sig}}_{Z\gamma}$     &  $471 \pm 28$   &  $578  \pm 29$ &  $347\pm 22$   &  $456  \pm 27$   \\ 
\hline
\hline
        Region  &\multicolumn{2}{c}{ $E_{\mathrm{T}}^\gamma> 60$ \GeV{}}&\multicolumn{2}{c}{$E_{\mathrm{T}}^\gamma> 60$ \GeV{}} \\
           &\multicolumn{2}{c}{$N_{\mathrm{jet}}\geq0$ }&\multicolumn{2}{c}{$N_{\mathrm{jet}}=0$} \\
       \hline
       $N^{\mathrm{obs}}_{Z\gamma}$ &   40            &  46               & 24    &  32    \\ 
       \hline
       $N^{\mathrm{BG}}_{Z\gamma}$              &  $4.1 \pm 2.4$  &  $5.1 \pm 3.3$    &  $1.6  \pm 1.6$     &  $2.1  \pm 2.1$    \\ 
       \hline
       $N^{\mathrm{sig}}_{Z\gamma}$     &  $35.9 \pm 6.7$ &  $40.9 \pm 7.1$   &  $22.4 \pm 5.1$   &  $29.9 \pm 5.9$   \\ 
       \hline
       \hline
\end{tabular}
\caption{Expected numbers of background events ($N^{BG}_{Z\gamma}$), observed numbers of signal events
    ($N^{sig}_{Z\gamma}$) and total numbers of events passing the selection
    requirements in the data ($N^{obs}_{Z\gamma}$) for the $pp \rightarrow e^{+}e^{-} \gamma$ channel and the $pp \rightarrow \mu^+\mu^- \gamma$ channel in different $E_{\mathrm{T}}^\gamma$ and jet multiplicity regions. The combined statistical and systematic uncertainties are shown. The uncertainty on the background prediction is dominated by systematic uncertainties in all regions. The background comes predominantly from $Z+$jets events.}
  \label{tab:zgbgsummary}
\end{table}


\section{Cross-section measurements}
\label{sec:cs}

The cross sections of the $W\gamma$ and $Z\gamma$ processes are measured as a function of the
photon $E_{\mathrm{T}}^\gamma$ threshold.
The measurements are performed in the fiducial region, defined at the particle level
using the objects and event kinematic selection criteria
described in Section~\ref{sec:Event_Selection}, and then extrapolated to an extended
fiducial region (as defined in Table~\ref{tab:fiducialcut}) common to the electron and muon final states.
Particle level is the simulation stage where stable particles, with lifetimes exceeding 10 ps,
are produced from the hard scattering or after hadronization, but before interacting with the detector.
The extrapolation is performed to correct for the signal acceptance loss in the calorimeter
transition region ($1.37<|\eta|<1.52$) for electrons and photons,
for the loss in the high $\eta$ region ($2.4<|\eta|<2.47$) for muons,
for the loss due to the $Z$-veto requirement in the $W\gamma$ electron channel,
and for the loss due to the transverse mass selection criteria in the $W\gamma$ analysis.
Jets at the particle level are reconstructed in MC-generated events by applying
the anti-$k_t$ jet reconstruction algorithm with a radius parameter $R$= 0.4
to all final state stable particles. To account for the effect of final state QED radiation,
the energy of the generated lepton at the particle level is defined as the energy of the
lepton after radiation plus the energy of all radiated photons within $\Delta R<0.1$ around the lepton direction.
Isolated photons with $\epsilon_h^p<0.5$ are considered as signal, where $\epsilon_h^p$ is defined
at particle level as the ratio between the sum of the energies carried by final state particles
in a cone $\Delta R<0.4$ around the photon direction and the energy carried by the photon.


%
%
%

\begin{table}[!htbp]
  \centering 
  \begin{tabular}{lcccc}
    \hline \hline
Cuts   &  $pp \rightarrow l\nu\gamma$  & $pp \rightarrow l^+ l^- \gamma$\\
\hline
Lepton  & $p_{\mathrm{T}}^l >25\GeV{}$ & $p_{\mathrm{T}}^l >25\GeV{}$\\
                        & $p_{\mathrm{T}}^\nu>25\ \GeV{}$    &  \\
  & $|\eta_l|<2.47$  & $|\eta_l|<2.47$  \\
\hline
Boson &            &   $m_{l^+l^-}>40$ \GeV{}  \\
\hline
Photon 
 & \multicolumn{2}{c}{{ Low $E_{\mathrm{T}}^\gamma$}: $E_{\mathrm{T}}^\gamma>15$  \GeV{}}\\
 & \multicolumn{2}{c}{{ Medium $E_{\mathrm{T}}^\gamma$}: $E_{\mathrm{T}}^\gamma>60$  \GeV{}}\\
 & \multicolumn{2}{c}{{ High $E_{\mathrm{T}}^\gamma$}: $E_{\mathrm{T}}^\gamma>100$  \GeV{}}& \\
& \multicolumn{2}{c}{$|\eta^\gamma|<2.37$, $\Delta R(l,\gamma)>0.7$ }\\
& \multicolumn{2}{c}{ photon isolation fraction $\epsilon_h^p < 0.5$}\\
\hline

Jet & \multicolumn{2}{c}{  $E_{\mathrm{T}}^{\mathrm{jet}}>30$ \GeV{}, $|\eta^{\mathrm{jet}}|<4.4$ }\\
 & \multicolumn{2}{c}{$\Delta R(e/\mu/\gamma,\mathrm{jet})>0.6$ }\\
 & \multicolumn{2}{c}{ Inclusive : $N^{\mathrm{jet}}\geq0$, Exclusive : $N^{\mathrm{jet}}=0$}\\

\hline

\hline \hline
  \end{tabular} 
  \caption{Definition of the extended fiducial region where the cross sections are
  evaluated; $p_{\mathrm{T}}^\nu$ is the transverse momentum of the neutrino from $W$ decays.}
\label{tab:fiducialcut}
\end{table}

The measurements of  cross sections for the processes $pp \to
l\nu\gamma+X$ and $pp \to l^+l^-\gamma+X$ are expressed as

\begin{equation}
\sigma_{pp \rightarrow l\nu\gamma(l^{+}l^{-}\gamma)}^{\mathrm{ext-fid}} = \frac{N^{\mathrm{\mathrm{sig}}}_{W\gamma(Z\gamma)}} {A_{W \gamma(Z\gamma)}\cdot  C_{W \gamma(Z\gamma)} \cdot L}
\label{Equ:cs_fid}
\end{equation}
where 
\begin{itemize}
\item $N_{W\gamma}^{\mathrm{\mathrm{sig}}}$ and $N_{Z\gamma}^{\mathrm{sig}}$ denote the numbers
  of background-subtracted signal events passing the selection
  criteria of the analyses in the $W \gamma$ and $Z \gamma$ channels.
  These numbers are listed in Table~\ref{tab:wgbg} and Table~\ref{tab:zgbgsummary}.
\item $L$ denotes the integrated
  luminosities for the channels of interest (1.02 fb$^{-1}$).
\item $C_{W\gamma}$ and $C_{Z\gamma}$ denote the ratios of the number of generated events which pass the final selection requirements after reconstruction to the number of generated events at particle level found within the fiducial region~\cite{WZpaper}.
\item $A_{W\gamma}$ and $A_{Z\gamma}$ denote the acceptances, defined at particle level as the ratio of the number of generated events found within the fiducial region to the number of generated events within the extended fiducial region.
\end{itemize}

The correction factors $C_{W\gamma}$ and $C_{Z\gamma}$ are shown in Table~\ref{tab:inputCA}. They are determined using the $W/Z+\gamma$ signal MC events and corrected with scale factors to account for small discrepancies between data and simulation. The uncertainties on $C_{W\gamma}$ and $C_{Z\gamma}$ due to the object selection efficiency are described in Section~\ref{sec:eff}. The uncertainties on $C_{W\gamma}$ and $C_{Z\gamma}$ due to the energy scale and resolution of the objects are summarized below.

The muon momentum scale and resolution are studied by comparing the invariant mass distribution of $Z\to \mu^+\mu^-$ events in data and MC simulation~\cite{WZpaper}. The uncertainty in the acceptance of the $W\gamma$ or $Z\gamma$ signal events due to the uncertainties in the muon momentum scale and resolution is $< 1\%$. Similarly the uncertainty due to the uncertainties in the EM energy scale and resolution is found to be $< 2.5\%$. The uncertainty from the jet energy scale and resolution on the exclusive $W\gamma$ and $Z\gamma$ signal acceptance varies in the range $5\% - 7\%$.  The uncertainty due to the $E_{\mathrm{T}}^{\mathrm{miss}}$ requirement is estimated to be $3\%$. It is due to several factors, including the uncertainty on the energy scale of the clusters reconstructed in the calorimeter that are not associated with any identified objects, and uncertainties from pile-up and muon momentum correction.

The overall relative uncertainties in $C_{W\gamma}$ and $C_{Z\gamma}$ are as large as 12.5\% in the low $E_{\mathrm{T}}^\gamma$ fiducial region and as large as 8.3\% in the medium and high $E_{\mathrm{T}}^\gamma$ fiducial region. They are dominated by the photon identification efficiency and the jet energy scale.

The acceptances $A_{W\gamma}$ and $A_{Z\gamma}$ are calculated using the signal MC simulation and shown in Table~\ref{tab:inputCA}. The systematic uncertainties are dominated by the limited knowledge of the PDFs ($<$1\%) and of the renormalization and factorization scales ($<$1\% for low $E_{\mathrm{T}}^\gamma$ region, $<$3.5\%  for medium and high $E_{\mathrm{T}}^\gamma$ region).


\begin{table}
  \centering 
  \begin{tabular}{cccccccccc}
    \hline \hline
$E_{\mathrm{T}}^\gamma $   & $>15$ \GeV{}   & $>60$ \GeV{}  & $>100$ \GeV{}   \\
\hline
\hline
   & \multicolumn{3}{c}{ $N_{\mathrm{jet}}=0$, $e$ channel}  \\
    \hline
$C_{W\gamma}$   & 0.402 $\pm$ 0.049    & 0.574 $\pm$ 0.045      & 0.517 $\pm$ 0.043   \\
$A_{W\gamma}$   & 0.762 $\pm$ 0.006     &  0.685 $\pm$ 0.017    &  0.672 $\pm$ 0.019 \\
\hline
$C_{Z\gamma}$   & 0.397 $\pm$ 0.045    & 0.592 $\pm$ 0.044     &  -  \\
$A_{Z\gamma}$   & 0.829 $\pm$ 0.014    &  0.834 $\pm$ 0.008   & -  \\
\hline
\hline
   & \multicolumn{3}{c}{ $N_{\mathrm{jet}}=0$, $\mu$ channel}  \\
    \hline
$C_{W\gamma}$   & 0.453 $\pm$ 0.054     & 0.653 $\pm$ 0.057      & 0.675 $\pm$ 0.059   \\
$A_{W\gamma}$   & 0.908 $\pm$ 0.006     &  0.764 $\pm$ 0.019    &  0.708 $\pm$ 0.017  \\
    \hline
$C_{Z\gamma}$  & 0.459 $\pm$ 0.052    & 0.641 $\pm$ 0.044     & -   \\
$A_{Z\gamma}$  & 0.915 $\pm$ 0.016    & 0.917 $\pm$ 0.008    &  -   \\
    \hline
\hline
   & \multicolumn{3}{c}{ $N_{\mathrm{jet}}\geq0$, $e$ channel}  \\
    \hline

$C_{W\gamma}$   & 0.453 $\pm$ 0.053    & 0.598 $\pm$ 0.036     & 0.576 $\pm$ 0.035  \\
$A_{W\gamma}$   & 0.725 $\pm$ 0.050     & 0.657 $\pm$ 0.011    &  0.666 $\pm$ 0.017 \\
\hline
$C_{Z\gamma}$   & 0.421 $\pm$ 0.044    & 0.609 $\pm$ 0.036     &  -  \\
$A_{Z\gamma}$   & 0.826 $\pm$ 0.014     &0.836 $\pm$ 0.050    & -  \\
\hline
\hline
   & \multicolumn{3}{c}{ $N_{\mathrm{jet}}\geq0$, $\mu$ channel}  \\
\hline
$C_{W\gamma}$ & 0.511 $\pm$ 0.057    & 0.650 $\pm$ 0.035       & 0.624 $\pm$ 0.035 \\
$A_{W\gamma}$ & 0.872 $\pm$ 0.005     &  0.776 $\pm$ 0.019     &  0.747 $\pm$ 0.023   \\
\hline
$C_{Z\gamma}$ & 0.485 $\pm$ 0.055    & 0.645 $\pm$ 0.035      & -   \\
$A_{Z\gamma}$ & 0.915 $\pm$ 0.016    & 0.917 $\pm$ 0.005     &  -   \\
    \hline \hline
  \end{tabular} 
  \caption{Summary of acceptance $A_{W\gamma}$($A_{Z\gamma}$) and correction factors $C_{W\gamma}$($C_{Z\gamma}$) for the calculation of the $W\gamma$($Z\gamma$) production cross sections. }
  \label{tab:inputCA} 
\end{table}

Assuming lepton universality for the $W$ and $Z$ boson decays, the measured cross sections in
the two channels are combined to reduce the statistical uncertainty.
For the combination, it is assumed that the uncertainties on the
lepton trigger and identification efficiencies are uncorrelated.
All other uncertainties, such as the uncertainties in the photon efficiency,
background estimation, and jet energy scale, are assumed to be fully correlated.
The measured production cross sections for the
$pp \rightarrow l\nu \gamma +X$ and $pp \rightarrow l^+l^- \gamma +X$
processes are summarized in Table~\ref{tab:cs}.


\section{Comparison with theoretical predictions}
\label{sec:theory}

\begin{table}
  \centering 
  \begin{tabular}{cccccccccc}
    \hline
    \hline
   & \multicolumn{1}{c}{$\sigma^{\mathrm{ext-fid}}$[pb]} & \multicolumn{1}{c}{  $\sigma^{\mathrm{ext-fid}}$[pb]}  \\
    \hline
    \hline
   & \multicolumn{1}{c}{ $E_{\mathrm{T}}^\gamma> 15$ \GeV{} } & \multicolumn{1}{c}{  $E_{\mathrm{T}}^\gamma> 15$ \GeV{}}  \\
   & \multicolumn{1}{c}{ exclusive} & \multicolumn{1}{c}{ inclusive}  \\
    \hline
$e\nu\gamma$   & 3.42  $\pm$ 0.14  $\pm$ 0.50          &4.35  $\pm$ 0.16   $\pm$ 0.64   \\
$\mu\nu\gamma$ &  3.23  $\pm$ 0.14  $\pm$ 0.48     &4.82  $\pm$ 0.15  $\pm$ 0.64   \\ 
$l\nu\gamma$    &  3.32  $\pm$ 0.10  $\pm$ 0.48     & 4.60  $\pm$ 0.11  $\pm$ 0.64  \\ 
    \hline
$ e^+e^-\gamma$   &  1.03  $\pm$ 0.06  $\pm$ 0.13     &1.32 $\pm$ 0.07  $\pm$ 0.16  \\ 
$ \mu^+\mu^-\gamma$  &  1.06  $\pm$ 0.05  $\pm$ 0.12     & 1.27  $\pm$ 0.06  $\pm$ 0.15 \\ 
$ l^+l^-\gamma$       &  1.05  $\pm$ 0.04  $\pm$ 0.12     & 1.29  $\pm$ 0.05  $\pm$ 0.15  \\
    \hline
    \hline
   & \multicolumn{1}{c}{ $E_{\mathrm{T}}^\gamma> 60$ \GeV{} } & \multicolumn{1}{c}{  $E_{\mathrm{T}}^\gamma> 60$ \GeV{}}  \\
   & \multicolumn{1}{c}{ exclusive} & \multicolumn{1}{c}{ inclusive}  \\
    \hline
$e\nu\gamma$    &  0.14  $\pm$ 0.02  $\pm$ 0.02     & 0.36  $\pm$ 0.03  $\pm$ 0.03 \\
$\mu\nu\gamma$  &  0.15  $\pm$ 0.02  $\pm$ 0.02     & 0.41  $\pm$ 0.03  $\pm$ 0.03  \\
$l\nu\gamma$   &  0.15  $\pm$ 0.01  $\pm$ 0.02    & 0.38  $\pm$ 0.02  $\pm$ 0.03  \\
    \hline
$ e^+e^-\gamma$   &  0.044  $\pm$ 0.010  $\pm$ 0.004     & 0.069  $\pm$ 0.012  $\pm$ 0.006\\
$ \mu^+\mu^-\gamma$   &  0.050  $\pm$ 0.010  $\pm$ 0.004     & 0.068  $\pm$ 0.011  $\pm$ 0.005  \\
$ l^+l^-\gamma$      &  0.047  $\pm$ 0.007  $\pm$ 0.004     & 0.068  $\pm$ 0.008  $\pm$ 0.005 \\
    \hline
    \hline
   & \multicolumn{1}{c}{ $E_{\mathrm{T}}^\gamma>$ 100 \GeV{} } & \multicolumn{1}{c}{  $E_{\mathrm{T}}^\gamma>$ 100 \GeV{}}  \\
   & \multicolumn{1}{c}{ exclusive} & \multicolumn{1}{c}{ inclusive}  \\

    \hline
$ e\nu\gamma$     &  0.040  $\pm$ 0.011  $\pm$ 0.009     & 0.114  $\pm$ 0.018  $\pm$ 0.010  \\
$ \mu\nu\gamma$   &  0.026  $\pm$ 0.008  $\pm$ 0.003     &0.135  $\pm$ 0.018  $\pm$ 0.010  \\
$ l\nu\gamma$      &  0.030  $\pm$ 0.006  $\pm$ 0.006      &  0.125  $\pm$ 0.013  $\pm$ 0.010 \\
    \hline \hline
  \end{tabular} 
  \caption{Measured cross sections for the $pp\to l\nu\gamma+X$ and $pp\to ll\gamma+X$ processes at $\sqrt{s}=7 \TeV{}$ in the extended fiducial region defined in Table ~\ref{tab:fiducialcut}. The first uncertainty is statistical and the second is systematic. The 3.7\% luminosity uncertainty is not included.}
  \label{tab:cs} 
\end{table}

\begin{table}
  \centering 
  \begin{tabular}{llcccc}
    \hline                        
    \hline                        
                              Channel &  $E_{\mathrm{T}}^\gamma$ (\GeV{}) &Cross section &Cross section \\
  &  &exclusive &  inclusive   \\
 \hline
 \hline
$pp\to l^{\pm}\nu\gamma$ &  $>$ 15 & 2.84$\pm$0.20 pb & 3.70$\pm$0.28 pb \\
   &   & (2.61$\pm$0.16 pb) & (3.58$\pm$0.26 pb) \\

$pp\to l^{\pm}\nu\gamma$& $>$ 60  & 134$\pm$21 fb & 260$\pm$38 fb\\
&   & (118$\pm$16 fb) & (255$\pm$35 fb)\\

$pp\to l^{\pm}\nu\gamma$ &  $>$ 100  & 34$\pm$5 fb& 82$\pm$13 fb \\
&     & (31$\pm$ 4 fb)& (80$\pm$12 fb)\\
$pp\to$ $l^+$ $l^-$$\gamma$ & $>$ 15   & 1.08$\pm$0.04 pb & 1.23$\pm$0.06 pb\\
                              &  & (1.03 $\pm$ 0.04 pb) & (1.22$\pm$0.05 pb)\\
$pp\to$ $l^+$ $l^-$$\gamma$ & $>$ 60  & 43$\pm$4 fb& 59$\pm$5 fb \\
                              &  & (40$\pm$3 fb) & (58$\pm$5 fb) \\
 \hline
    \hline                        
 \end{tabular} 
 \caption{ Expected NLO inclusive and exclusive cross sections for the $pp\to l^{\pm}\nu\gamma+X$ and $pp\to$ $l^+$ $l^-$$\gamma+X$ processes in the extended fiducial region as defined in Table ~\ref{tab:fiducialcut}. The cross sections are quoted at particle (parton) level as described in the text.  }

  \label{tab:MCFMcs} 
\end{table}

The {\sc mcfm}~\cite{MCFM} program is used to predict the NLO cross section for $pp$ $\to$ $l^{\pm}\nu\gamma+X$ and $pp$ $\to$ $l^{+}$$l^{-}$$\gamma+X$ production. It includes photons from direct $W\gamma$ and $Z\gamma$ diboson production, from final state radiation off the leptons in the $W/Z$ decays and from quark/gluon fragmentation into an isolated photon. Possible effects of composite $W$ and $Z$ boson structure can be simulated through the introduction of aTGCs. Event generation is done using the MSTW2008NLO~\cite{pdfmstw} parton distribution functions and the default electroweak parameters of {\sc mcfm}. The kinematic requirements for the parton-level generation are the same as those chosen at particle level for the extended fiducial cross-section measurements (see Table~\ref{tab:fiducialcut}).
The resulting parton-level SM predictions for the cross sections are summarized by the numbers in parentheses in Table~\ref{tab:MCFMcs}. 
These are quoted as \emph{inclusive}, using only the lepton and photon selection cuts, and \emph{exclusive}, requiring no quark/gluon with $|$$\eta$$|$ $<$ 4.4 and $E_{\mathrm{T}}>30$ \GeV{} in the final state. The cross-section uncertainties are dominated by the PDF uncertainty, the scale uncertainty and the uncertainty due to the photon isolation fraction. The scale uncertainty is evaluated by varying the renormalization and factorization scales by factors of 2 and 1/2 around the nominal scale $M_{W/Z}$. The PDF uncertainty is estimated using the MSTW2008NLO PDFs' error eigenvectors at their 90$\%$ confidence-level (CL) limits. The uncertainty due to photon isolation
fraction is evaluated by varying $\epsilon_h$ from 0.0 to 1.0. Here $\epsilon_h$ is defined at parton level as the ratio of the sum of the energies carried by the partons in the cone $\Delta R<0.4$ around the photon direction to the energy carried by the photon. The variation in the predicted cross section due to the choice of $\epsilon_h$ threshold is a conservative estimate of the uncertainty in matching the parton-level photon isolation to the photon isolation criteria applied in the experimental measurement. The total uncertainties in the $W\gamma$ ($Z\gamma$) NLO cross-section predictions are 7$\%$ (5$\%$) for photon $E_{\mathrm{T}}^\gamma>$ 15 GeV and 14$\%$ (8$\%$) for photon $E_{\mathrm{T}}^\gamma>$ 60 GeV.

To compare the SM cross-section predictions to the measured cross section, the theoretical predictions must be corrected for the difference between jets defined at the parton level (single quarks or gluons) and jets defined at the particle level as done for the cross-section measurement.
These corrections account for the difference in jet definitions and in photon isolation definitions between the particle level and the parton level. The {\sc alpgen}+{\sc herwig} (for $W\gamma$) and {\sc sherpa} (for $Z\gamma$) MC samples
are used to estimate these parton-to-particle scale factors $S_{W\gamma}$ and $S_{Z\gamma}$. They increase the parton-level cross sections by typically $5\%$ with uncertainties that vary from 2$\%$ to 9$\%$ depending on the channel. These uncertainties are evaluated from the differences in the $S_{W\gamma}$ and $S_{Z\gamma}$ values obtained from several generators. 

\begin{figure*}
  \centering
\subfigure[]{\includegraphics[width=0.99\columnwidth]{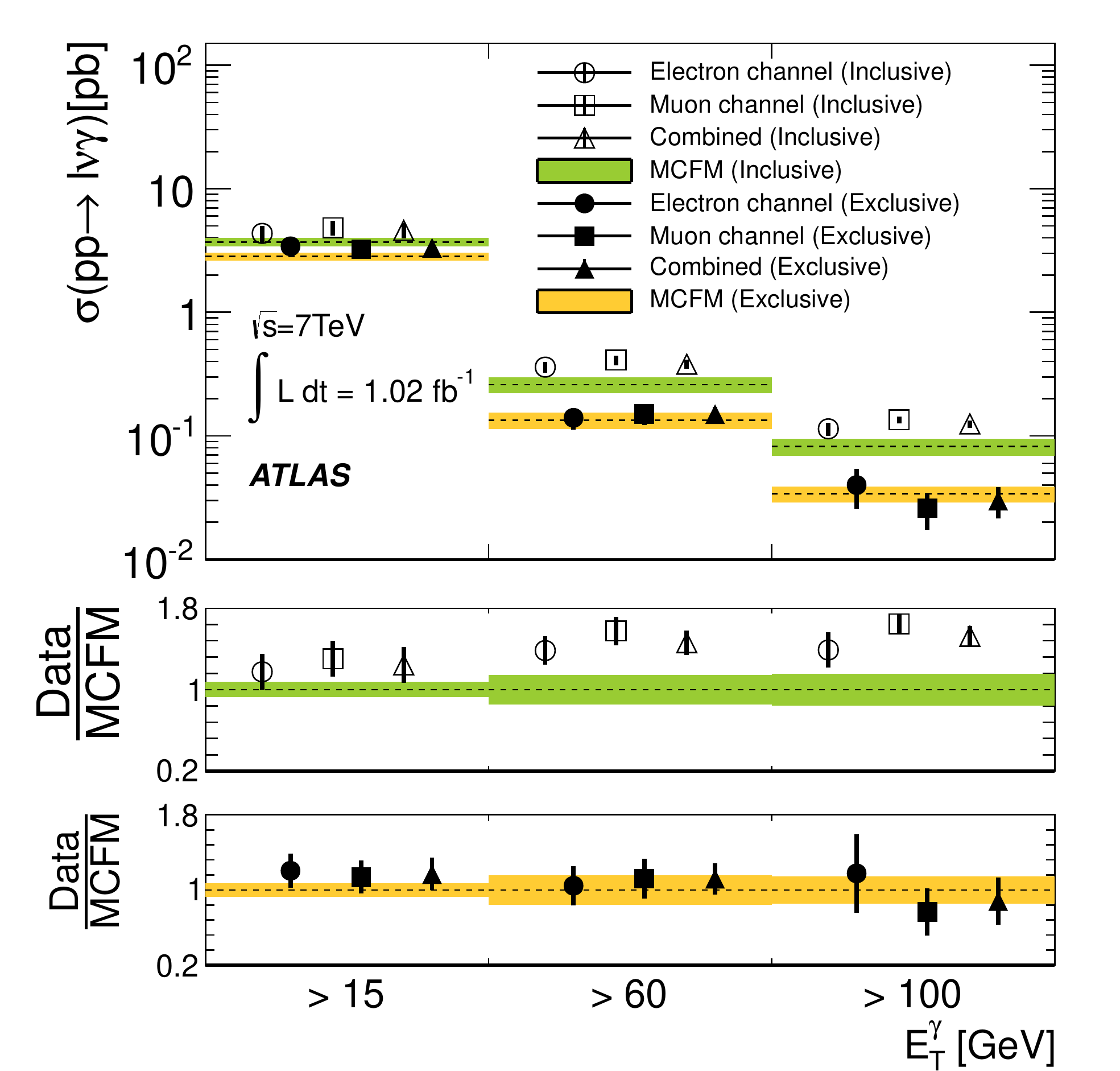}}
\subfigure[]{\includegraphics[width=0.99\columnwidth]{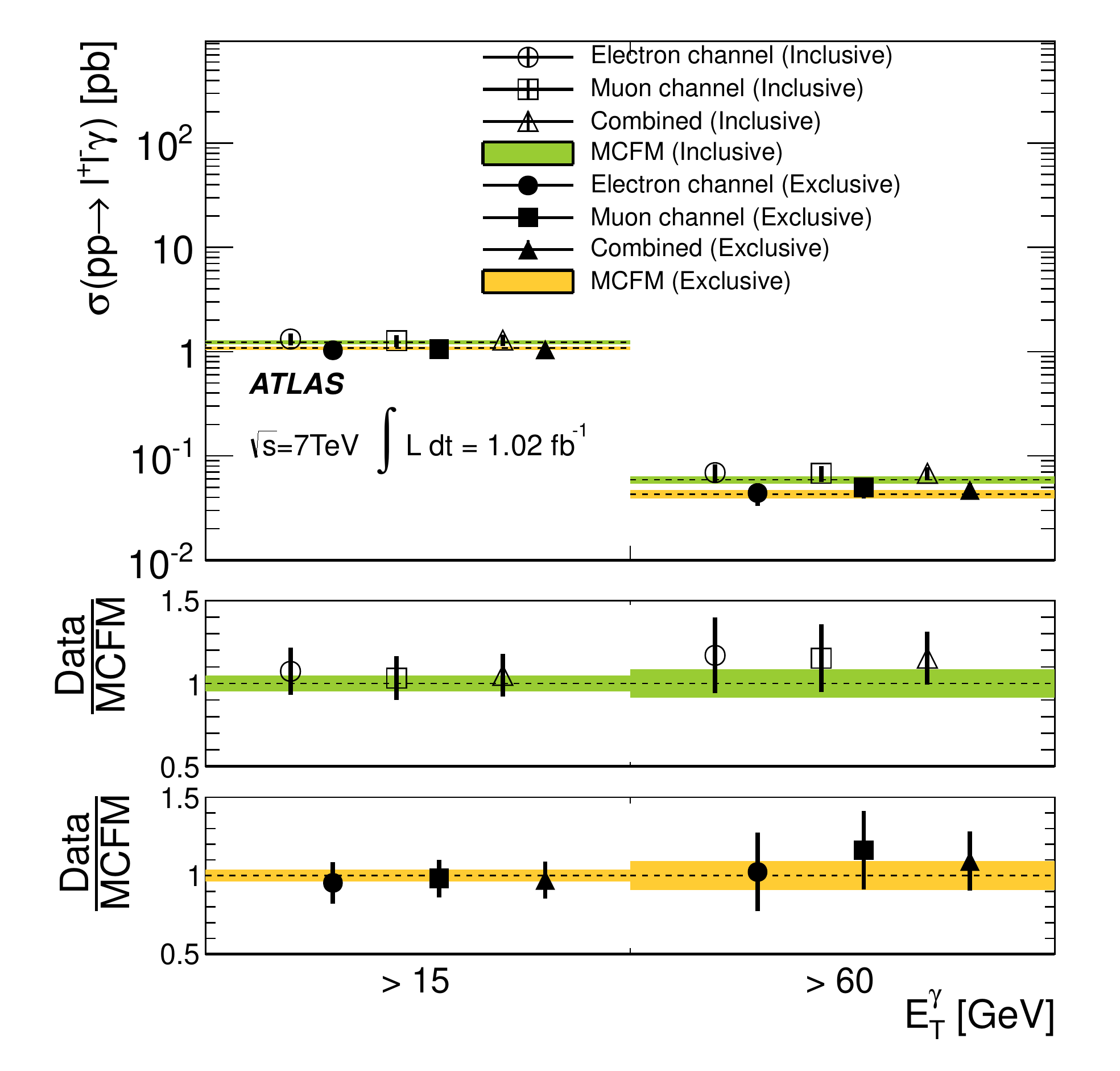}}
  \caption{The measured cross section for (a) $W\gamma$ production, (b) $Z\gamma$ production as a function of the photon transverse energy, in the extended fiducial region as defined in Table~\ref{tab:fiducialcut}, together with the SM model prediction. The lower plots show the ratio between the data and the prediction of the MCFM generator.}
  \label{fig:totalcswzg}
\end{figure*}

The SM predictions for the particle-level (parton-level) cross sections are summarized in Table~\ref{tab:MCFMcs}. The uncertainties quoted include those from the {\sc mcfm} parton-level generator predictions, photon isolation matching to the data, and the scaling from parton to particle-level cross sections. 
Figure~\ref{fig:totalcswzg} presents a summary of all cross-section measurements of $W\gamma$ and $Z\gamma$ production made in this study and the corresponding particle-level SM expectations. There is good agreement between the measured cross sections for the exclusive events and the {\sc mcfm} prediction. 

 For inclusive production, the {\sc mcfm} NLO cross-section prediction includes real parton emission processes only up to one radiated quark or gluon. 
The lack of higher-order QCD contributions results in an underestimation of the predicted cross sections, especially for events with high $E_{\mathrm{T}}^\gamma$ photons, which have significant contributions from multi-jet final states.
 Figure~\ref{fig:Wg_kin_elec3} shows that the multi-jet contribution is important in the $W\gamma$ processes. Therefore higher-order jet production is needed in the MC simulation (see Section ~\ref{sec:simulation}) to describe the photon transverse energy spectrum with the inclusive selection and the jet multiplicity distribution in $W\gamma$ and $Z\gamma$ events, as shown in Figure~\ref{fig:Wg_kin_elec1} and Figure~\ref{fig:Wg_kin_elec3}.

\section{Limits on  Anomalous Triple Gauge Couplings}
\label{sec:ATGC}

The spectra of high energy photons in $W\gamma$ and $Z\gamma$ events
are sensitive to new phenomena that alter the couplings among the gauge bosons.
These effects can be described by modifying the $WW\gamma$ coupling $\kappa_\gamma$
 from its SM value of one and adding terms with new couplings to the
$WW\gamma$ and $ZV\gamma$ ($V=\gamma$ or $Z$) interaction Lagrangian.
Assuming C and P conservation separately, the aTGC parameters are
generally chosen as $\lambda_{\gamma}$ and $\Delta \kappa_{\gamma}$ ($\Delta \kappa_{\gamma} = \kappa_{\gamma} -1$) for the
$WW\gamma$ vertex~\cite{BaurLO,BaurNLO}, 
and $h^{V}_{3}$ and $h^{V}_{4}$ for the $ZV\gamma$ vertices~\cite{PhysRevD.47.4889}.
Form factors are introduced to avoid unitarity violation at very high energy.
Typical choices of these form factors for the $WW\gamma$  aTGCs are:
$\Delta\kappa_\gamma(s) = \Delta\kappa_{\gamma}/(1+\hat{s}/\Lambda^2)^2$ and
$\lambda_\gamma(s) = \lambda_{\gamma}/(1+\hat{s}/\Lambda^2)^2$~\cite{BaurNLO}.
For the $ZV\gamma$ aTGCs, conventional choices of form factors are
$h^{V}_{3}(s) =  h^{V}_{3}/(1+\hat{s}/\Lambda^2)^3$ and
$h^{V}_{4}(s) = h^{V}_{4}/(1+\hat{s}/\Lambda^2)^4$~\cite{PhysRevD.47.4889}.
Here $\sqrt{\hat{s}}$ is the $W\gamma$ or $Z\gamma$ invariant mass
and $\Lambda$ is the new physics energy scale.
Deviations of the aTGC parameters from the SM predictions of zero lead to an excess of
high energy photons associated with the $W$ and $Z$ bosons.
 
\begin{figure*}
  \centering

  \subfigure[]{\includegraphics[width=0.99\columnwidth]{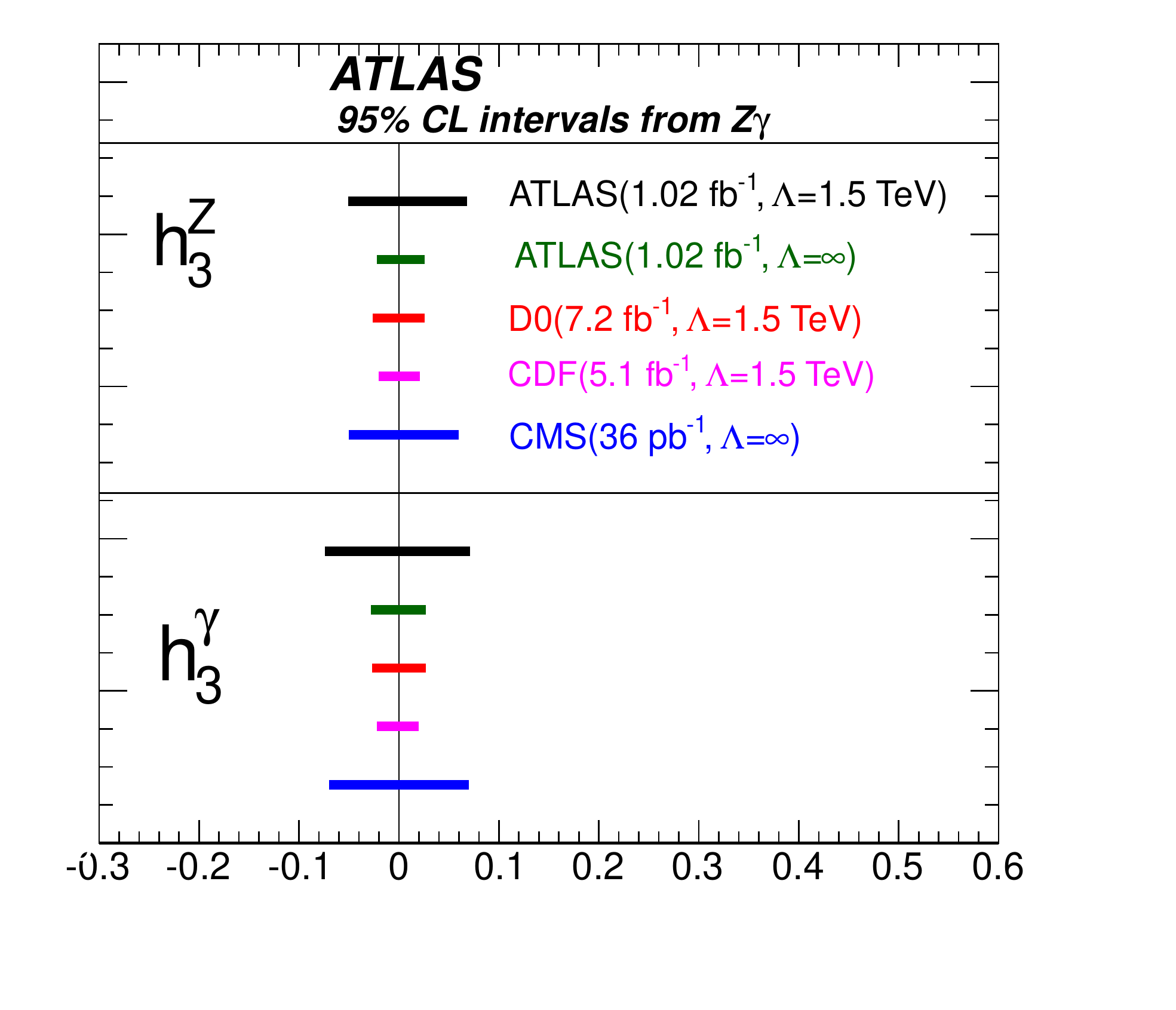}}
  \subfigure[]{\includegraphics[width=0.99\columnwidth]{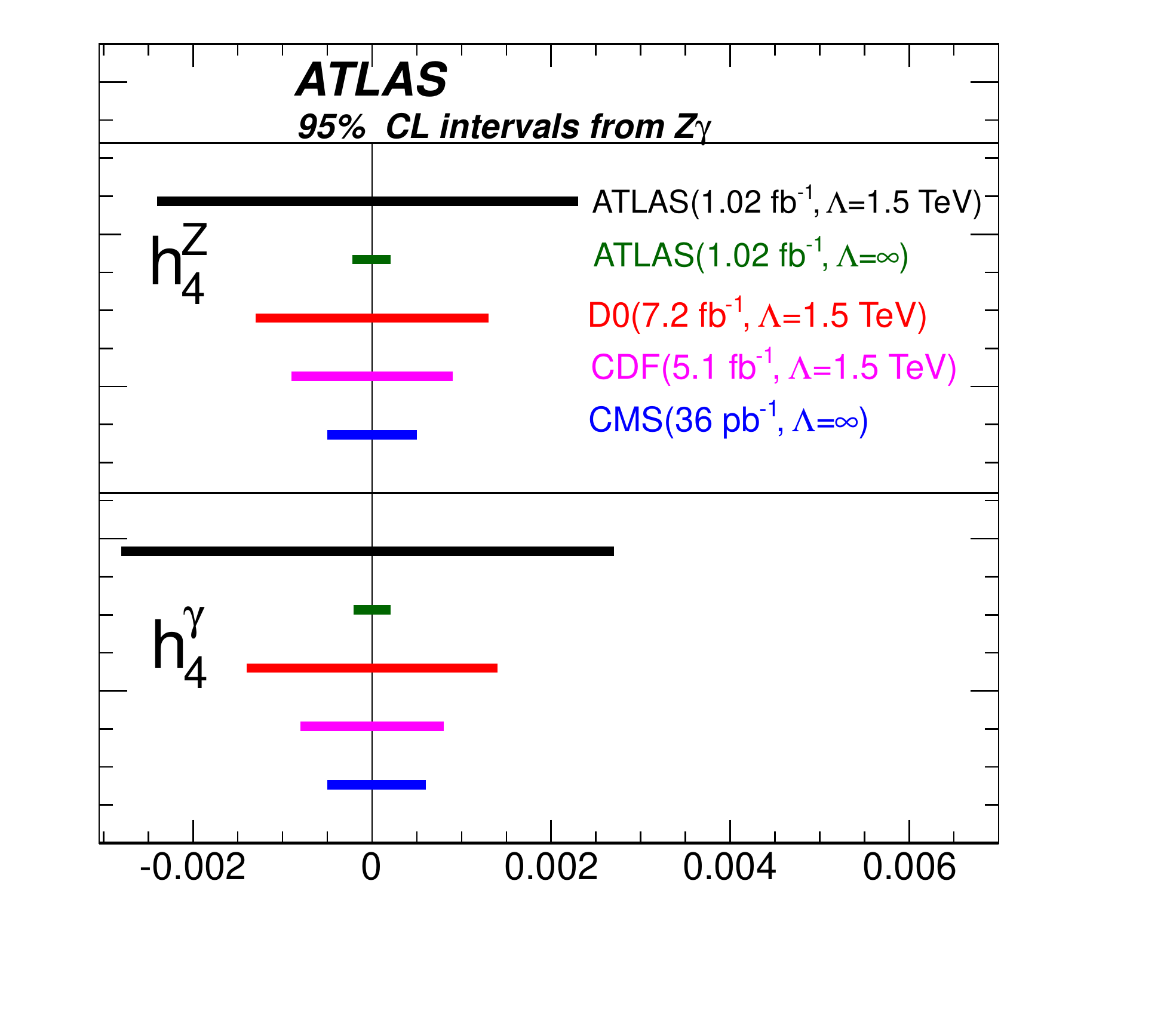}}
  \subfigure[]{\includegraphics[width=0.99\columnwidth]{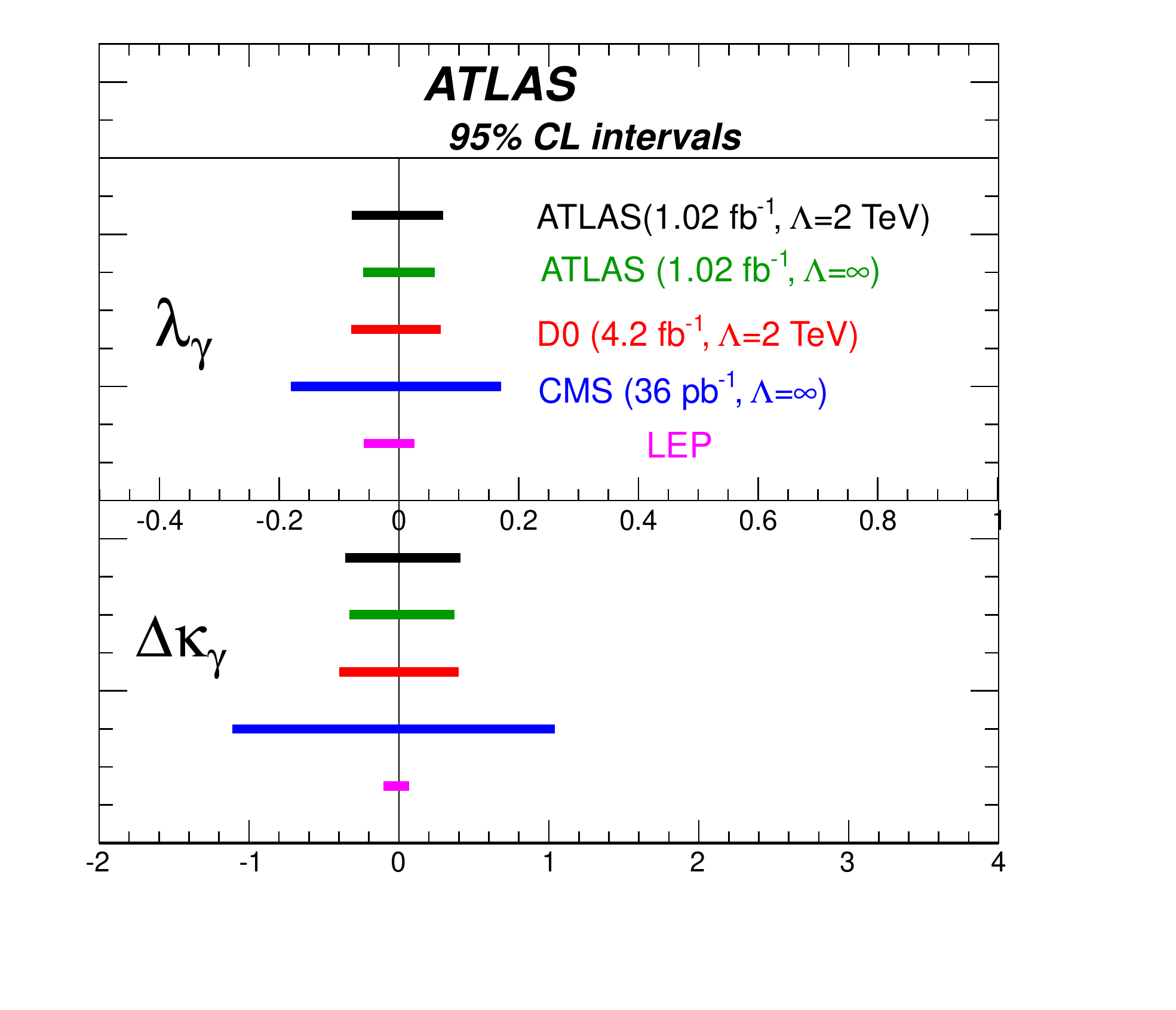}}
  \caption{ The 95\% CL intervals for anomalous couplings from ATLAS, D0~\cite{D0paper2},
           CDF~\cite{CDFpaper}, CMS~\cite{CMSpaper} and LEP~\cite{LEP} for (a),(b) the neutral aTGCs $h_{3}^{\gamma}$, $h_{3}^{Z}$, $h_{4}^{\gamma}$, $h_{4}^{Z}$ as obtained from $Z\gamma$ events, and (c) the charged aTGCs $\Delta\kappa_{\gamma}$, $\lambda_{\gamma}$. Integrated luminosities and new physics scale parameter $\Lambda$ are shown. The ATLAS, CMS and Tevatron results for the charged aTGCs are measured from $W\gamma$ production. The LEP charged aTGC results are obtained from $WW$ production, which is sensitive also to the $WWZ$ couplings and hence required some assumptions about the relations between the $WW\gamma$ and $WWZ$ aTGCs~\cite{LEP,LEP1,LEP2,LEP3}. The sensitivity of the LEP data to neutral aTGCs is much smaller than that of the hadron colliders; therefore the LEP results have not been included in (a) and (b). }
  \label{fig:limit}
\end{figure*}

\begin{table}[h]
\begin{center}
 \begin{tabular}{cccc}
\hline
\hline
%
                 & Measured & Measured    & Expected \\ \hline
  $\Lambda$          & 2 \TeV{} & $\infty$    & $\infty$  \\\hline
 $\Delta\kappa_{\gamma}$ & (-0.36,0.41)  & (-0.33,0.37)       &   (-0.33,0.36)     \\
$\lambda_{\gamma}$      & (-0.079,0.074) &   (-0.060,0.060)     &   (-0.063,0.055)  \\ \hline
  $\Lambda$          & 1.5 \TeV{}  & $\infty$   & $\infty$  \\\hline
  $h_{3}^{\gamma}$ & (-0.074,0.071) &   (-0.028,0.027)    &   (-0.027,0.027)    \\
  $h_{3}^{Z}     $ & (-0.051,0.068)   &   (-0.022,0.026)     &   (-0.022,0.025)    \\
  $h_{4}^{\gamma}$ & (-0.0028,0.0027) &   (-0.00021,0.00021)   &   (-0.00021,0.00021)\\
  $h_{4}^{Z}$      & (-0.0024,0.0023) &   (-0.00022,0.00021)    & (-0.00022,0.00021) \\ \hline\hline
\end{tabular}
 \caption[]{The measured and expected 95\% CL intervals on the charged
($\Delta\kappa_{\gamma}$, $\lambda_{\gamma}$) and
neutral ($h_{3}^{\gamma}$, $h_{3}^{Z}$, $h_{4}^{\gamma}$, $h_{4}^{Z}$) anomalous couplings.
The results obtained using different $\Lambda$ values are shown. The two numbers in each
parentheses denote the 95\% CL interval.}
 \label{Tab:wg_atgc}\vspace{12pt}
\end{center}
\end{table}

Measurements of the exclusive extended fiducial cross sections for $W\gamma$ production with $E_{\mathrm{T}}^\gamma>100$ \GeV{} and $Z\gamma$ production with $E_{\mathrm{T}}^\gamma>60$ \GeV{} are used to extract aTGC limits. The cross-section predictions with aTGCs ($\sigma^{\mathrm{aTGC}}_{W\gamma}$ and $\sigma^{\mathrm{aTGC}}_{Z\gamma}$) are obtained from the {\sc mcfm} generator.
The number of expected $W\gamma$ events in the exclusive extended fiducial region
($N^{\mathrm{aTGC}}_{W\gamma}(\Delta\kappa_\gamma, \lambda_\gamma)$) for given aTGCs are obtained as
$N^{\mathrm{aTGC}}_{W\gamma}(\Delta\kappa_\gamma, \lambda_\gamma) = \sigma^{\mathrm{aTGC}}_{W\gamma} \times C_{W\gamma}  \times A_{W\gamma} \times S_{W\gamma} \times L$.
For the $Z\gamma$ case, $N^{\mathrm{aTGC}}_{Z\gamma}(h_3^\gamma,h_4^\gamma)$ or $N^{\mathrm{aTGC}}_{Z\gamma}(h_3^Z,h_4^Z)$ are obtained in a similar way.
The anomalous couplings influence the kinematic properties of $W\gamma$ and $Z\gamma$ events and thus the corrections for event reconstruction ($C_{W\gamma}$ and $C_{Z\gamma}$). The maximum variations of $C_{W\gamma}$ and $C_{Z\gamma}$ within the measured aTGC limits are quoted as additional systematic uncertainties.
The limits on a given aTGC parameter (e.g. $h_{i}^{V}$) are extracted from the Bayesian posterior, given the extended fiducial measurements.
The Bayesian posterior probability density function is obtained by integrating over the nuisance parameters corresponding to all systematic uncertainties and assuming a flat Bayesian prior in $h_{i}^{V}$.
This calculation has been done for multiple values of the scale parameter $\Lambda$ in order
to be able to compare these results with those from LEP~\cite{LEP}, Tevatron~\cite{CDFpaper, D0paper, D0paper2} and CMS~\cite{CMSpaper}.
The limits are defined as the values of aTGC parameters which demarcate the central 95\% of the
integral of the likelihood distribution. The resulting allowed ranges for the anomalous 
couplings are shown in Table~\ref{Tab:wg_atgc} for $WW\gamma$ and $ZV\gamma$. 
The results are also shown in Figure~\ref{fig:limit}, along with the LEP, Tevatron and CMS measurements.


\section{Summary}
\label{sec:summary}

The production of $W\gamma$ and $Z\gamma$ boson pairs in 7 TeV{}
$pp$ collisions has been studied using 1.02 fb$^{-1}$ of data
collected with the ATLAS detector. The measurements have been made using the
$pp \to l^{\pm}\nu\gamma + X$ and $pp \to l^+ l^-\gamma +
X$ final states, where the charged lepton is an electron or muon and
the photons are required to be isolated.
%
The results are compared to SM predictions using a NLO parton-level generator. The NLO SM predictions for the exclusive $W\gamma$ and $Z\gamma$ production cross sections agree well with the data for events with both low (15 \GeV{}) and high (60 \GeV{} or 100 \GeV{}) photon $E_{\mathrm{T}}^\gamma$ thresholds. For the high photon thresholds, where multi-jet production dominates, the measured inclusive $W\gamma$ cross sections are higher than the NLO calculations for the inclusive $pp \to l^{\pm}\nu\gamma +X$ process, which do not include multiple quark/gluon emission.
 The measurements are also compared to LO MC generators with multiple quark/gluon emission in the matrix element calculations. These LO MC predictions reproduce the shape of the photon $E_{\mathrm{T}}^\gamma$ spectrum and the kinematic properties of the leptons and jets in the $W\gamma$ and $Z\gamma$ candidate events.


The measurements of exclusive $W\gamma$ ($Z\gamma$) production with $E_{\mathrm{T}}^\gamma >$ 100 (60) \GeV{} are used to constrain anomalous triple gauge couplings ($\lambda_{\gamma}$, $\Delta$$\kappa_{\gamma}$, $h^{V}_{3}$ and $h^{V}_{4}$). No evidence for physics beyond the SM is observed. The limits obtained in this study are compatible with those from LEP and Tevatron and are more stringent than previous LHC results.

\section*{Acknowledgements}
We thank CERN for the very successful operation of the LHC, as well as the
support staff from our institutions without whom ATLAS could not be
operated efficiently.
We also thank John Campbell, Keith Ellis and Ciaran Williams for their advice
about theory calculations using the {\sc MCFM} program. 

We acknowledge the support of ANPCyT, Argentina; YerPhI, Armenia; ARC,
Australia; BMWF, Austria; ANAS, Azerbaijan; SSTC, Belarus; CNPq and FAPESP,
Brazil; NSERC, NRC and CFI, Canada; CERN; CONICYT, Chile; CAS, MOST and NSFC,
China; COLCIENCIAS, Colombia; MSMT CR, MPO CR and VSC CR, Czech Republic;
DNRF, DNSRC and Lundbeck Foundation, Denmark; EPLANET and ERC, European Union;
IN2P3-CNRS, CEA-DSM/IRFU, France; GNAS, Georgia; BMBF, DFG, HGF, MPG and AvH
Foundation, Germany; GSRT, Greece; ISF, MINERVA, GIF, DIP and Benoziyo Center,
Israel; INFN, Italy; MEXT and JSPS, Japan; CNRST, Morocco; FOM and NWO,
Netherlands; RCN, Norway; MNiSW, Poland; GRICES and FCT, Portugal; MERYS
(MECTS), Romania; MES of Russia and ROSATOM, Russian Federation; JINR; MSTD,
Serbia; MSSR, Slovakia; ARRS and MVZT, Slovenia; DST/NRF, South Africa;
MICINN, Spain; SRC and Wallenberg Foundation, Sweden; SER, SNSF and Cantons of
Bern and Geneva, Switzerland; NSC, Taiwan; TAEK, Turkey; STFC, the Royal
Society and Leverhulme Trust, United Kingdom; DOE and NSF, United States of
America.

The crucial computing support from all WLCG partners is acknowledged
gratefully, in particular from CERN and the ATLAS Tier-1 facilities at
TRIUMF (Canada), NDGF (Denmark, Norway, Sweden), CC-IN2P3 (France),
KIT/GridKA (Germany), INFN-CNAF (Italy), NL-T1 (Netherlands), PIC (Spain),
ASGC (Taiwan), RAL (UK) and BNL (USA) and in the Tier-2 facilities
worldwide.

\bibliographystyle{atlasnote}
\bibliography{plb_wzgamma2011}
\onecolumngrid
\clearpage 
\begin{flushleft}
{\Large The ATLAS Collaboration}

\bigskip

G.~Aad$^{\rm 48}$,
B.~Abbott$^{\rm 111}$,
J.~Abdallah$^{\rm 11}$,
S.~Abdel~Khalek$^{\rm 115}$,
A.A.~Abdelalim$^{\rm 49}$,
O.~Abdinov$^{\rm 10}$,
B.~Abi$^{\rm 112}$,
M.~Abolins$^{\rm 88}$,
O.S.~AbouZeid$^{\rm 158}$,
H.~Abramowicz$^{\rm 153}$
% ...
\bigskip

% ...
$^{10}$ Institute of Physics, Azerbaijan Academy of Sciences, Baku, Azerbaijan\\
$^{11}$ Institut de F\'isica d'Altes Energies and Departament de F\'isica de la Universitat Aut\`onoma  de Barcelona and ICREA, Barcelona, Spain\\
% ...
$^{48}$ Fakult\"{a}t f\"{u}r Mathematik und Physik, Albert-Ludwigs-Universit\"{a}t, Freiburg i.Br., Germany\\
$^{49}$ Section de Physique, Universit\'e de Gen\`eve, Geneva, Switzerland\\
% ...
$^{88}$ Department of Physics and Astronomy, Michigan State University, East Lansing MI, United States of America\\
% ...
$^{111}$ Homer L. Dodge Department of Physics and Astronomy, University of Oklahoma, Norman OK, United States of America\\
$^{112}$ Department of Physics, Oklahoma State University, Stillwater OK, United States of America\\
% ...
$^{115}$ LAL, Universit\'e Paris-Sud and CNRS/IN2P3, Orsay, France\\
% ...
$^{153}$ Raymond and Beverly Sackler School of Physics and Astronomy, Tel Aviv University, Tel Aviv, Israel\\
% ...
$^{158}$ Department of Physics, University of Toronto, Toronto ON, Canada\\
% ...
\end{flushleft}


\end{document}